%% file: main.tex
\documentclass[journal]{IEEEtran}
\ifCLASSINFOpdf

\else

\fi

\usepackage{amsmath}
\usepackage{amsfonts}
\usepackage{hyperref}
\usepackage{cleveref}
\usepackage{cite}
\usepackage{xcolor}
\usepackage{graphicx} 
\usepackage{subfigure}
\usepackage{array} 
\usepackage{multirow} 

\begin{document}

\title{BSM-iMagLS: ILD Informed Binaural Signal Matching for Reproduction with Head-Mounted Microphone Arrays}

\author{Or~Berebi,~\IEEEmembership{Student Member,~IEEE,}
        Zamir~Ben-Hur,~\IEEEmembership{Member,~IEEE,}
        David~Lou~Alon,~\IEEEmembership{Member,~IEEE,}
        and Boaz~Rafaely,~\IEEEmembership{Senior Member,~IEEE}
\thanks{O. Berebi and B. Rafaely are with the School of Electrical and Computer Engineering, Ben-Gurion University of the Negev, Beer-Sheva 84105, Israel.}
\thanks{Z. Ben-Hur and D. Lou~Alon are with Reality Labs Research, Meta, 1 Hacker Way, Menlo Park, CA 94025, USA.}}

\maketitle

\begin{abstract}
Headphone listening in applications such as augmented and virtual reality (AR and VR) relies on high-quality spatial audio to ensure immersion, making accurate binaural reproduction a critical component. As capture devices, wearable arrays with only a few microphones with irregular arrangement face challenges in achieving a reproduction quality comparable to that of arrays with a large number of microphones. Binaural signal matching (BSM) has recently been presented as a signal-independent  approach for generating high-quality binaural signal using only a few microphones, which is further improved using magnitude-least squares (MagLS) optimization at high frequencies. This paper extends BSM with MagLS by introducing interaural level difference (ILD) into the MagLS, integrated into BSM (BSM-iMagLS). Using a deep neural network (DNN)-based solver, BSM-iMagLS achieves joint optimization of magnitude, ILD, and magnitude derivatives, improving spatial fidelity. Performance is validated through theoretical analysis, numerical simulations with diverse HRTFs and head-mounted array geometries, and listening experiments, demonstrating a substantial reduction in ILD errors while maintaining comparable magnitude accuracy to state-of-the-art solutions. The results highlight the potential of BSM-iMagLS to enhance binaural reproduction for wearable and portable devices.
\end{abstract}


\IEEEpeerreviewmaketitle

\input{sections/Ch1}
\input{sections/Ch2}

\input{sections/Ch3}

\input{sections/Ch4_v2}
\input{sections/Ch5_v5}

\input{sections/Ch6_v3}
\input{sections/Ch7}

\bibliographystyle{./bibtex/IEEEtran}
\bibliography{./bibtex/refs}

\input{bios}

\end{document}

%% file: sections/Ch1.tex
\section{Introduction}


\IEEEPARstart{B}{inaural} audio reproduction aims to recreate the acoustic conditions of natural spatial hearing by rendering the ear-specific signals that a listener would receive in a real acoustic environment, and delivering them over headphones.This approach is widely used in spatial audio applications such as virtual and augmented reality, teleconferencing, and advanced hearing aids~\cite{begault20003, vorlander2015virtual, rafaely2022spatial}.

A common solution to spatial audio reproduction involves the use of higher-order Ambisonics (HOA). The HOA signals are typically captured by a spherical microphone array and transformed into the spherical harmonic (SH) domain using the spherical Fourier transform. These SH-domain signals are then filtered using head-related transfer functions (HRTFs) that have been similarly represented in the SH domain. This filtering step, involving a weighted sum across the SH coefficients, yields the binaural signals at the ears~\cite{zotter2019ambisonics}. However, despite its maturity and incorporation into standards like MPEG-H Audio~\cite{herre2015mpeg}, HOA requires specific array geometries such as spherical arrays, and high spatial resolution, limiting its suitability for wearable and portable devices~\cite{zotter2019ambisonics}.

Parametric spatial audio approaches offer an alternative with flexibility in array geometry. These rely on estimating parameters of the sound field, such as direction of arrival of dominant sound sources, and diffuseness of the reverberant components, and use these estimates within a signal model to derive binaural signals~\cite{politis2018compass, mccormack2022parametric,hold2023optimizing}. While parametric methods can achieve high fidelity when the model assumptions hold, their performance is sensitive to estimation errors and relies on assumptions like sparse source distributions in the time-frequency domain. This signal dependency limits their robustness in complex acoustic environments. Conversely, signal-independent methods that do not require detailed sound field information could be desirable when only limited information about the recorded sound field is available.

Signal-independent beamforming-based approaches, such as the recently proposed Binaural Signal Matching (BSM), are designed to reproduce binaural signals from arbitrary microphone arrays. It operates by determining a set of linear filters that, when applied to the microphone signals, approximate the desired binaural output. This approach provides flexibility for arbitrary array geometries and does not require prior information about the captured scene~\cite{madmoni2020beamforming, madmoni2025design}. Previous works demonstrated BSM's potential while also highlighting its limited accuracy at high frequencies and degraded spatial fidelity, especially under head movements. In these works, the binaural reproduction error was shown to be related to the degradation of binaural cues such as the ILD.

The problem of poor ILD accuracy in binaural reproduction based on Ambisonics signals and microphone array recordings has been previously identified, particularly in relation to SH-based methods. Strategies for optimizing ILD in binaural reproduction from Ambisonics signals have been proposed to address spatial perception issues~\cite{mckenzie2019interaural}. Techniques for magnitude-corrected and time-aligned HRTF interpolation have also been developed to improve localization and spectral accuracy, and shown to improve ILD~\cite{arend2023magnitude}. Furthermore, the encoding of equatorial microphone arrays into Ambisonics has been studied, providing further insight into how array geometry influences ILD~\cite{ahrens2022ambisonic}. Although these approaches focus on spherical harmonic representations, they underscore the importance of accurately preserving ILD cues.

Building on prior work, two key steps were previously introduced to improve signal-independent binaural rendering with microphone arrays. In~\cite{berebi2023imagls}, an ILD-informed magnitude least-squares (iMagLS) formulation was proposed for spherical microphone arrays. The method was defined and analyzed in the SH domain, optimizing first-order Ambisonics SH coefficients of HRTFs under magnitude and ILD constraints. In~\cite{berebi2024feasibility}, a feasibility study extended this concept to arbitrary array geometries by introducing the idea of iMagLS optimization of BSM coefficients. However, this short-form conference paper offered only preliminary numerical results and did not present a formal optimization framework or subjective evaluation. The current manuscript extends and generalizes these previous efforts. The specific contributions of this work are:

\begin{enumerate}
\item \textbf{Improved formulation:} A new BSM-iMagLS formulation is introduced, adding \textit{first-derivative magnitude matching} to the previously suggested magnitude and ILD loss terms.
\item \textbf{DNN-based optimization framework:} A deep neural network (DNN)-based optimization framework is proposed to efficiently solve the BSM-iMagLS objective function, offering a scalable and differentiable alternative to classical solvers.
\item \textbf{Extended theoretical analysis:} A detailed analysis is provided of the limitations and accuracy bounds for ILD matching under arbitrary microphone geometries.
\item \textbf{Comprehensive numerical validation:} The method is evaluated using both simulated and measured array transfer functions, as well as real-world HRTF datasets. In addition, a head rotation compensation analysis is performed to further validate the robustness of the approach, thereby extending the limited numerical analysis in~\cite{berebi2024feasibility}.
\item \textbf{Subjective listening evaluation:} A head-tracked listening experiment is conducted to assess perceptual differences between BSM-iMagLS and prior BSM methods, which were not previously evaluated subjectively.
\end{enumerate}

%% file: sections/Ch2.tex
\section{Mathematical Background}
This section presents fundamental concepts related to the recorded signal model, the binaural signals model, and the BSM method.
\subsection{Signal Model}
In this context, it is assumed that an $M$-element microphone array is available, with its central positioning at the origin of a spherical coordinate system. The array records the pressure variations in a sound field containing $Q$ far-field sources, each generating a signal at the origin denoted as $\{ s_q(k) \}_{q=1}^{Q}$ and associated with their respective directions-of-arrival (DOA) as $\{ \Omega_q = (\theta_q,\phi_q) \}_{q=1}^{Q}$. The wave number, $k$, is defined as $k=\frac{2 \pi}{\lambda}$, where $\lambda$ represents the wavelength. The pressure measured by the array is modeled within the narrow-band framework as follows~\cite{van2002optimum}:
\begin{equation}\label{mic_model}
    \mathbf{x}(k) = \mathbf{V}(k)\mathbf{s}(k) + \mathbf{n}(k) 
\end{equation}
Here, $\mathbf{x}(k) = [ \, x_1(k), \dots,x_M(k) ] ^T$ denotes a vector of length $M$ representing the microphone signals, with $(.)^T$ denoting the transpose operation. The matrix $\mathbf{V}(k) = [ \, \mathbf{v}_1(k),\dots, \mathbf{v}_Q(k) ] $ is an $M \times Q$ matrix, where the $q$-th column contains the steering vector for the $q$-th source. The collection of $Q$ steering vectors can be written as $ \{ \mathbf{v}_q(k) = [ \, v(k,\mathbf{d}_1 ; \Omega_q), \dots,v(k,\mathbf{d}_M ; \Omega_q) ] ^T \}_{q=1}^{Q}$, where $\mathbf{d}_m$ represents the Cartesian coordinates of the $m$-th microphone in the array, and $v(k,\mathbf{d}_m ; \Omega_q)$ signifies the transfer function between the far-field source at DOA $\Omega_q$ and the microphone at $\mathbf{d}_m$. This transfer function can be determined analytically, numerically, or through measurements for various array configurations~\cite{van2002optimum,rafaely2015fundamentals}. For $q=1,\dots,Q$, $\mathbf{s}(k) = [ \, s_1(k), \dots,s_Q(k) ] ^T$ denotes the source signals vector with a length of $Q$, and $\mathbf{n}(k) = [ \, n_1(k), \dots,n_M(k) ] ^T$ represents an additive noise vector of length $M$.

\subsection{Binaural Signals Model}\label{ch2:BSM}
Assuming, as previously mentioned, that the sound field consists of $Q$ far-field sources, the calculation of left and right ear pressures for a listener positioned at the microphone array location is achieved through the use of the listener's HRTF:
\begin{equation}\label{true_binaural_signals}
    p^{l,r}(k) = [ \mathbf{h}^{l,r}(k) ] ^T \mathbf{s}(k)
\end{equation}
Here, $\mathbf{h}^{l,r}(k) = [ \, h^{l,r}(k,\Omega_1), \dots,h^{l,r}(k,\Omega_Q) ] ^T$, representing a $Q$-length vector, describes the transfer function between the far-field source at DOA $\Omega_q$ and the listener's left or right ear, denoted by $(.)^l$ and $(.)^r$, respectively. The narrow-band signals presented in this section can be transformed into the time domain, making them suitable for playback over headphones for binaural reproduction.

\subsection{Binaural Signal Matching (BSM)}\label{sec:BSM}
The left and right ear binaural signals can be calculated from the measurements in Eq.~\eqref{mic_model} using BSM as detailed in~\cite{madmoni2025design}:
\begin{equation}\label{BSM_rendering}
    z^{l,r} = [\mathbf{c}^{l,r}]^H\mathbf{x}
\end{equation}

In this context, $\mathbf{c}^{l,r} = [ \, c^{l,r}_1, \dots,c^{l,r}_M ] ^T$, where $(.)^H$ denotes the Hermitian operation, signifies the filter coefficients associated with the left and right ears. Note that the wave number index $k$ has been omitted for brevity, however, the derivation remains within the narrow-band domain. 

The calculation of the coefficients $\mathbf{c}^{l,r}$ is accomplished by minimizing a dissimilarity measure individually for each ear, involving a comparison between the true binaural signal, as outlined in Eq.~\eqref{true_binaural_signals}, and its BSM approximation, represented by Eq.~\eqref{BSM_rendering}:
\begin{equation}\label{dissimilarity_measure}
    \epsilon^{l,r} = D(p^{l,r}, z^{l,r})
\end{equation}
The computation of $\mathbf{c}^{l,r}$ involves solving the minimization problem presented as follows:
\begin{equation}\label{min_problem}
    \mathbf{c}^{l,r}_{BSM} = \arg \min_{\mathbf{c}^{l,r}} \epsilon^{l,r}
\end{equation}
The choice of $D(.,.)$ can significantly influence the spectral and spatial quality of $z^{l,r}$, as demonstrated in~\cite{madmoni2025design}. The study considered both Mean Square Error (MSE) 
\begin{equation}\label{D_MSE}
    D_{MSE}(p^{l,r},z^{l,r})=E[|p^{l,r}-z^{l,r}|^2]
\end{equation}
where $E[.]$ denotes the expectation operation, and Magnitude Square Error (MagLS)
\begin{equation}\label{D_MagLS}
    D_{MLS}(p^{l,r},z^{l,r})=E[|\, |p^{l,r}|-|z^{l,r}| \,|^2].
\end{equation}

To address Eq.~\eqref{min_problem} for $D_{MSE}$, we start by assuming that the source signals $\{ s_q(k) \}_{q=1}^{Q}$ and the noise components $\{ n_m(k) \}_{m=1}^{M}$ are uncorrelated. Subsequently, by substituting~\eqref{true_binaural_signals} and~\eqref{BSM_rendering} into~\eqref{D_MSE}, the expression results in:
\begin{equation}\label{D_MSE_explicit_error}
\begin{split}
    D_{MSE} & = \left( [\mathbf{c}^{l,r}]^H\mathbf{V} - [\mathbf{h}^{l,r}]^T \right)\mathbf{R}_s \left( [\mathbf{c}^{l,r}]^H\mathbf{V} - [\mathbf{h}^{l,r}]^T \right)^H \\ 
    & + [\mathbf{c}^{l,r}]^H \mathbf{R}_n[\mathbf{c}^{l,r}]
\end{split}
\end{equation}
Here, $\mathbf{R}_s = E[\mathbf{s}\mathbf{s}^H]$ and $\mathbf{R}_n = E[\mathbf{n}\mathbf{n}^H]$. This study concentrates on scenarios where minimal information about the sound field is accessible. Consequently, the sound sources are assumed to be uncorrelated, or originating from a diffuse field, while the noise is assumed uncorrelated between microphones, or spatially white, e.g. sensor noise. In such instances, the correlation matrices are simplified to $\mathbf{R}_s =\sigma_s^2 \mathbf{I}_Q$ and $\mathbf{R}_n =\sigma_n^2 \mathbf{I}_M$, where $\sigma_s^2$ and $\sigma_n^2$ denote the source and noise variance, and $\mathbf{I}_Q$ and $\mathbf{I}_M$ are identity matrices of size $Q$ and $M$, respectively. This simplification allows for further simplification of Eq.~\eqref{D_MSE_explicit_error} to:
\begin{equation}\label{D_MSE_diffused}
    D_{MSE} = \sigma_s^2 \| \mathbf{V}^T [\mathbf{c}^{l,r}]^* - \mathbf{h}^{l,r} \|_2^2 + \sigma_n^2 \| [\mathbf{c}^{l,r}]^* \|_2^2
\end{equation}
Here, $\| .\|_2^2$ signifies the $l^2$ norm. Finally, The solution to Eq.~\eqref{min_problem} for~\eqref{D_MSE_diffused}, as outlined in~\cite{madmoni2025design}, is expressed as:
\begin{equation}\label{BSM-MSE}
    \mathbf{c}^{l,r}_{MSE} = \left(\mathbf{V}\mathbf{V}^H + \frac{\sigma_n^2}{\sigma_s^2}\mathbf{I}_M\right)^{-1} \mathbf{V} [\mathbf{h}^{l,r}]^*
\end{equation}
The solution to Eq.~\eqref{BSM-MSE} will be referred to as BSM-Least Squares (BSM-LS) throughout this study.

The solution presented in~\eqref{BSM-MSE} is derived under the assumption that the sources and noise signals are both uncorrelated and white. Additionally, it assumes that the sound field consists of a finite set of sources with known DOAs. However, in various scenarios, these assumptions may not be valid, or information about the sources may be unavailable. Despite this, previous work~\cite{madmoni2025design, madmoni2025design} demonstrated that accurate binaural signals can be reproduced even in complex acoustic environments and with arbitrary DOAs using the solution $\mathbf{c}^{l,r}_{MSE}$. Specifically, $\mathbf{c}^{l,r}_{MSE}$ are designed to perform well in a diffuse-field scenario—characterized by a large number of independent sources with uniformly distributed DOAs—which supports its signal-independent nature. As long as the overall error remains sufficiently low, $\mathbf{c}^{l,r}_{MSE}$ can yield accurate binaural signals regardless of the specific source configuration~\cite{madmoni2025design}.

The performance of BSM-LS is found to be dependent on the number of microphones $M$. In situations where $M$ is relatively small, Eq.~\eqref{BSM-MSE} might yield a relatively large error, particularly in the high-frequency range. This is primarily attributed to the increase in the effective spherical harmonics (SH) order of $\mathbf{h}^{l,r}$~\cite{zhang2010insights}. As a result, it has been demonstrated that solving Eq.~\eqref{min_problem} for $D_{MLS}$ in the high-frequency range can lead to an enhancement in performance~\cite{madmoni2025design}.

Solving for $D_{MLS}$ involves matching absolute values instead of complex values, and the choice of $D_{MLS}$ is based on the notion that inter-aural level differences (ILD) are more crucial than inter-aural time differences (ITD) for spatial perception at high frequencies~\cite{klumpp1956some,zwislocki1956just,brughera2013human,macpherson2002listener}. Unlike $D_{MSE}$, where an analytic global minimum solution is available due to the convex nature of the MSE problem, the MagLS problem is non-convex, making a global minimum potentially unattainable. Fortunately, various approaches for finding a local minimum for Eq.~\eqref{min_problem} with $D_{MLS}$ exist, involving iterative numerical solvers~\cite{kassakian2006convex}. In this study, we employ an algorithm previously suggested in~\cite{zotter2019ambisonics} for Ambisonics binaural rendering and utilized in~\cite{mccormack2023six} for BSM rendering to minimize $D_{MLS}$. The MagLS coefficients are computed as follows:
\begin{equation}\label{BSM-MLS}
    \mathbf{c}^{l,r}_{MLS} = \left(\mathbf{V}\mathbf{V}^H + \frac{\sigma_n^2}{\sigma_s^2}\mathbf{I}_M\right)^{-1} \mathbf{V} [\mathbf{\hat{h}}^{l,r}]^*
\end{equation}
In this formulation, a modified version of the original HRTF is utilized, which incorporates only its magnitude information along with an added phase component. This modification can be expressed as:
\begin{equation}\label{H_hat}
    \mathbf{\hat{h}}^{l,r}(f) =
    \begin{cases}
    \mathbf{h}^{l,r}(f) &,\quad   f < f_c \\
    |\mathbf{h}^{l,r}(f)| e^{i \mathbf{\Phi}^{l,r}(f)} &,\quad f \ge f_c
    \end{cases}
\end{equation}
As indicated in~\cite{mccormack2023six}, the phase component is defined as the phase response of the reconstructed HRTFs for the preceding frequency index, given by
\begin{equation}\label{eq:mls_phase}
    \mathbf{\Phi}^{l,r}(f_n) = \angle{\left[  [\mathbf{c}^{l,r}_{MLS}(f_{(n-1)})]^H \mathbf{V}(f_{(n-1)}) \right]}
\end{equation}
Here, $\angle[\cdot]$ denotes the element-wise phase value of the enclosed matrix entries. The cutoff frequency $f_c$ is typically set at $1.5$ kHz, a choice informed by duplex theory~\cite{strutt1907our}. Above this threshold, the solution aims to more effectively preserve magnitude, resulting in a significant reduction of magnitude error compared to the BSM-MSE solution.



%% file: sections/Ch3.tex
\section{Proposed Method BSM-iMagLS}\label{sec:3}

\begin{figure*}
  \includegraphics[width=\textwidth,height=1.4cm]{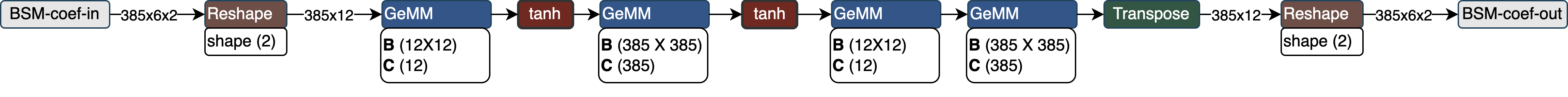}
  \caption{Network diagram example with $M=6$, $\text{NFFT}=385$, and $\tanh{(.)}$ used as the activation function. GeMM is an abbreviation of general matrix multiplication.}
  \label{fig1:DNN}
\end{figure*}

This section introduces the proposed dissimilarity measure for binaural reproduction using BSM-iMagLS. Furthermore, a framework for solving Eq.~\eqref{min_problem} using the proposed dissimilarity measure is outlined.

\subsection{BSM-iMagLS Dissimilarity Measure}
As mentioned in Section~\ref{sec:BSM}, the preference for $D_\text{MLS}$ of Eq.~\eqref{D_MagLS} over $D_\text{MSE}$ of Eq.~\eqref{D_MSE} for higher frequencies was motivated by perceptual considerations. The newly introduced $D_\text{iMLS}$, denoted as Inter-aural Level Differences with Magnitude Least Square (iMagLS), extends this rationale by incorporating a joint optimization of both ears and explicitly considering the level difference ratio. In this regard, $D_\text{iMLS}$ can be viewed as a natural extension of $D_\text{MLS}$, as it not only aims to match the absolute values of each ear independently but also to preserve the ratio between them. This extension may be particularly advantageous in frequency ranges where the error in absolute values is significant due to array constraints, while maintaining a low ILD error remains feasible. ILD is a critical auditory cue for sound localization, and inaccuracies in ILD can lead to perceptual shifts, as the human auditory system depends on these cues to identify the position of sound sources~\cite{katz2018binaural}.

The proposed dissimilarity measure $D_\text{iMLS}$ is expressed as follows, incorporating the original MagLS dissimilarity measure, and two new measures representing the derivatives of magnitude over frequency, and ILD dissimilarity:
\begin{equation} \label{D_iMLS}
\begin{split}
D_\text{iMLS}(p^{l,r}, z^{l,r}) & = (D_\text{MLS}(p^l,z^l)+D_\text{MLS}(p^r,z^r))  \\
 & + \lambda_1(D_\text{dMLS}(p^l,z^l)+D_\text{dMLS}(p^r,z^r)) \\
 & + \lambda_2 D_\text{ILD}(p^l,p^r, z^l,z^r)
\end{split}
\end{equation}
Here, the parameters $\lambda_{1,2} \in \mathbb{R}$ serves as a weighting factor for balancing the contributions of $D_\text{MLS}$, $D_\text{dMLS}$  and $D_\text{ILD}$ to the overall error. 

The ILD dissimilarity, denoted as $D_\text{ILD}$, is defined as:
\begin{equation}\label{ILD_error_eq}
    D_\text{ILD}(p^{l,r}, z^{l,r}) = \frac{1}{|F|} \sum_{f_i \in F} \left\| \text{ILD}_{\text{ref}}(\Omega,f_i) - \text{ILD}_{bsm}(\Omega,f_i) \right\|^2_2
\end{equation}
where, $\frac{1}{|F|} \sum_{f_i \in F}$ denotes the average over the set of Gammatone filters center frequencies $f_i$, and $\| \cdot \|_2$ represents the $2$-norm over directions $\Omega$. Furthermore, $\text{ILD}_{ref}(\Omega,f_i)$ and $\text{ILD}_{bsm}(\Omega,f_i)$ are defined as~\cite{xie2013head}:
\begin{equation}\label{ILD_eq_ref}
    \text{ILD}_{ref}(\Omega,f_i) = 10 \log_{10}\frac{\int_{f_1}^{f_2} G(f_i,f)|p^l(\Omega,f)|^2 df}{\int_{f_1}^{f_2} G(f_i,f)|p^r(\Omega,f)|^2 df} ,
\end{equation}

\begin{equation}\label{ILD_eq_BSM}
    \text{ILD}_{bsm}(\Omega,f_i) = 10 \log_{10}\frac{\int_{f_1}^{f_2} G(f_i,f)|z^l(\Omega,f)|^2 df}{\int_{f_1}^{f_2} G(f_i,f)|z^r(\Omega,f)|^2 df} .
\end{equation}
Here, $G(f_i, f)$ denotes the filter function centered at the frequency $f_i$, while $f_1$ and $f_2$ represent the frequency integration boundaries for ILD evaluation, typically set at $f_1 = 1.5$ kHz and $f_2 = 20$ kHz~\cite{xie2013head}. The evaluation of ILD is performed over the horizontal plane, specifically within the range of $\Omega \in (\theta = 90^\circ, 0^\circ \le \phi < 360^\circ)$ with $(\theta,\phi)$ denoting the elevation and azimuth angles in a spherical coordinate system. Notably, $\text{ILD}_{ref}(\Omega,f_i)$ uses $p^{l,r}$ from Eq.~\eqref{true_binaural_signals} with $\mathbf{s}(k)=1$ and $\Omega_q = \Omega, Q=1$, while $\text{ILD}_{bsm}(\Omega,f_i)$ utilizes $z^{l,r}$ from Eq.~\eqref{BSM_rendering} for similar signals and directions.

An extra term is incorporated into the overall dissimilarity measure of Eq.~\eqref{D_iMLS}, denoted as $ D_\text{dMLS}$, which accounts for matching the first derivative between $ |z^{l,r}| $ and $ |p^{l,r}| $. This can be expressed as:
\begin{equation}\label{D_dMagLS}
    D_{dMLS}(p^{l,r},z^{l,r}) = E\left[\left| \frac{d|p^{l,r}|}{df} - \frac{d|z^{l,r}|}{df} \right|^2 \right]
\end{equation}
This weighted term improves the first derivatives matching, leading to a smoother magnitude error curve. Statistical assumptions on $p^{l,r}$ and $z^{l,r}$ are identical to those described in Eq.~\eqref{D_MagLS}.
The inclusion of $D_{\text{dMLS}}$ was motivated by observations that minimizing $D_{\text{iMLS}}$ without $D_{\text{dMLS}}$ results in non-smooth magnitude error curves, which in turn introduced audible artifacts. The addition of this term helped mitigate these perceived spectral distortions.

While previous work~\cite{madmoni2025design} showed that accurate binaural reproduction is possible in complex acoustic environments using $\mathbf{c}^{l,r}_{MSE}$ under a diffuse-field assumption, it is not immediately clear that this extends to the MagLS and iMagLS. The motivation for using $\mathbf{c}^{l,r}_{MLS}$ and $\mathbf{c}^{l,r}_{iMLS}$, stems from the limited performance of $\mathbf{c}^{l,r}_{MSE}$ at high frequencies. Importantly, even when the MagLS design leads to low binaural error, phase distortions may persist—particularly at high frequencies. Since MagLS relies on the reduced perceptual sensitivity to phase errors, it is not guaranteed that accurate reproduction in complex environments can be achieved in the same way. Therefore, the effectiveness of $\mathbf{c}^{l,r}_{MLS}$ and $\mathbf{c}^{l,r}_{iMLS}$ under such conditions must be validated experimentally.

\subsection{BSM-iMagLS Formulation}
The proposed dissimilarity measure $D_\text{iMLS}$ can be integrated into the BSM framework by substituting it into Eq.~\eqref{dissimilarity_measure} and solving Eq.~\eqref{min_problem}. This leads to the optimization problem:
\begin{equation}\label{iMagLS_loss_eq}
    \mathbf{c}^{l,r}_{\text{iMLS}} = \arg \min_{\mathbf{c}^{l,r}} D_\text{iMLS}(p^l,p^r, z^l,z^r)
\end{equation}
However, similar to using $D_\text{MLS}$, this problem is non-convex, so finding a global minimum is not guaranteed. Nevertheless, a local minimum can be obtained through numerical methods. The following section details the approach to reach this local minimum.

Suppose we are given an HRTF $\mathbf{h}^{l,r}$ and the array steering matrix $\mathbf{V}$ for a set of angles $\{ \Omega_a \}_{a=1}^{A}$. Using Eq.~\eqref{BSM-MSE}, an initial set of BSM coefficients $\mathbf{c}^{l,r}$ can be computed. Additionally, $D_\text{iMLS}$ can also be calculated, as both $p^{l,r}$ from Eq.~\eqref{true_binaural_signals} and $z^{l,r}$ from Eq.~\eqref{BSM_rendering} can be derived when the Signal-to-Noise Ratio (SNR) is sufficiently high to neglect $\mathbf{n}$ in Eq.~\eqref{mic_model}.

A Deep Neural Network (DNN) is utilized to solve Eq.~\eqref{iMagLS_loss_eq} for the given $\mathbf{h}^{l,r}$ and $\mathbf{V}$. Its primary objective is to match the pair $\{ \mathbf{c}^{l,r}_{a}, p_{a}^{l,r} \}_{a=1}^{A}$ under the loss $D_\text{iMLS}$~\cite{deng2014deep}. Although other iterative solvers could be used to solve Eq.~\eqref{iMagLS_loss_eq} \cite{broyden1970convergence, berebi2024feasibility, berebi2023imagls}, experimentation shows that the proposed DNN-based method converges faster compared to these alternatives. The DNN mapping from some initial $\mathbf{c}^{l,r}$ to $\mathbf{c}^{l,r}_{\text{iMLS}}$ is represented as:
\begin{equation}\label{iMLS_loss1}
    \mathbf{c}^{l,r}_{\text{iMLS}} = f_{\boldsymbol{\beta}}(\mathbf{c}^{l,r})
\end{equation}
where $f_{\boldsymbol{\beta}}(.)$ is a DNN with $\boldsymbol{\beta}$ denoting the network parameters. To solve Eq.~\eqref{iMagLS_loss_eq}, the parameters of $f_{\boldsymbol{\beta}}(.)$ are adjusted so that $D_\text{iMLS}$ reaches a local minimum, or in other words, $D_\text{iMLS}$ serves as the loss function. Eq.~\eqref{iMagLS_loss_eq} can now be expressed in terms of $\boldsymbol{\beta}$:
\begin{equation}
    \hat{\boldsymbol{\beta}} = \arg \min_{\boldsymbol{\beta}} D_\text{iMLS}(f_{\boldsymbol{\beta}}(\mathbf{c}^{l,r}), \mathbf{V},\mathbf{h}^{l,r})
\end{equation}
where $p^{l,r}$ and $z^{l,r}$ used in Eq.~\eqref{D_iMLS} for $D_\text{iMLS}$ can be computed from $(f_{\boldsymbol{\beta}}(\mathbf{c}^{l,r}), \mathbf{V},\mathbf{h}^{l,r})$ and Eqs.~\ref{true_binaural_signals} and~\ref{BSM_rendering}. It is important to note that $f_{\boldsymbol{\beta}}(.)$ is not trained to provide a global solution for varying conditions in $\mathbf{h}^{l,r}$ and $\mathbf{V}$, but to solve each case individually.

\subsection{DNN architecture}\label{ch:3.3}
In this study, a feed-forward multi-layered perceptron (MLP) serves as the architecture for the DNN. The initial $\mathbf{c}^{l,r}$ tensor, with dimensions $[\text{NFFT}, M, 2]$, where $\text{NFFT}$ represents the frequency bins, is reshaped into $[\text{NFFT}, 2M]$. This reshaped tensor undergoes a linear layer with the activation function $\sigma(\cdot)$, defined as:
\begin{equation}
    \textbf{y}_1 = \sigma \left(   \mathbf{c}^{l,r}\mathbf{W}_1 +\mathbf{b}_1   \right)
\end{equation}
Here, $\mathbf{W}_1$ is a $2M \times 2M$ complex matrix that operates across the dimensions of microphones and ears. Subsequently, a second layer is applied on the frequency bins, given by:
\begin{equation}
    \textbf{y}_2 = \sigma \left(   \textbf{y}_1^T\mathbf{W}_2 +\mathbf{b}_2   \right)
\end{equation}
In this case, $\mathbf{W}_2$ and $\mathbf{b}_2$ have dimensions $\text{NFFT} \times \text{NFFT}$ and $2M \times \text{NFFT}$, respectively. Following this, another linear layer is applied to the $2M$ dimension, and then another linear layer to the $\text{NFFT}$ dimension:
\begin{equation}
    \textbf{y}_3 = (\textbf{y}_2^T \mathbf{W}_3 +\mathbf{b}_3)^T \mathbf{W}_4 +\mathbf{b}_4
\end{equation}
This series of operations is performed to produce output values that can exceed the typical range of $\tanh(\cdot)$, i.e., values greater than 1 or less than –1, thereby overcoming the inherent output limitations of the activation function used in this work, as shown in Fig.~\ref{fig1:DNN}. Finally, $\textbf{y}_3$ is reshaped to match the original dimension of $\mathbf{c}^{l,r}$:
\begin{equation}
    \mathbf{y} = \text{reshape}(\textbf{y}_3)
\end{equation}
The trainable parameters of the DNN are explicitly represented as $\boldsymbol{\beta}= [\textbf{W}_1,\textbf{W}_2,\textbf{W}_3,\textbf{W}_4,\textbf{b}_1,\textbf{b}_2,\textbf{b}_3,\textbf{b}_4,]$. An example of the DNN block diagram is shown in Fig.~\ref{fig1:DNN}.

The DNN is trained using the ADAM optimizer~\cite{kinga2015method} for $T$ iterations until a local minimum of $D_\text{iMLS}$ and $\hat{\boldsymbol{\beta}}$ is reached. The final iMLS BSM coefficients are given by:
\begin{equation}
    \mathbf{c}^{l,r}_{\text{iMLS}} = f_{\hat{\boldsymbol{\beta}}}(\mathbf{c}^{l,r})
\end{equation}

%% file: sections/Ch4_v2.tex
\section{Performance Limitations for ILD Error in BSM}\label{sec:4}
This section derives a theoretical formulation of the ILD error in relation to a reference for BSM rendering, with a focus on isolating the contributing variables to better understand the ILD performance limits and identify its key characteristics. In particular, we explore whether zero ILD error can be achieved and under what conditions, therefore leading to ideal ILD performance.

To simplify the analysis, we derive the narrow-band ILD error computed for each individual frequency bin for a plane wave arriving from direction $\Omega$. This differs from the Gammatone band ILD error defined in Eq.~\eqref{ILD_error_eq}, which is evaluated over perceptually motivated wide frequency bands. The analysis explores how to design BSM coefficients that achieve zero ILD error for a single direction, and proposes solutions for specific cases, such as when the number of directions matches the number of microphones. It also discusses the dependencies of the ILD error on the HRTF and the acoustic transfer function (ATF) of the array, highlighting the complexities involved in minimizing the error across multiple directions.

The expression for the narrow-band ILD error due to a single plane wave arriving from direction $\Omega$ is given by~\eqref{ILD_error_eq}:
\begin{align}\label{narrow_band_ILD_e}
    \epsilon_{ILD}(\Omega,f) & = \left|10\log{\left(\frac{|p^l(\Omega,f)|^2}{|p^r(\Omega,f)|^2}\right)} -  10\log{\left(\frac{|z^l(\Omega,f)|^2}{|z^r(\Omega,f)|^2}\right)} \right|   \nonumber\\
                & = \left|20\log{\left(\frac{|p^l|}{|p^r|}\frac{|z^r|}{|z^l|}\right)}\right|
\end{align}
Substituting the expressions for the binaural signals from the reference model (Eq.~\eqref{true_binaural_signals}) and the BSM rendering (Eq.~\eqref{BSM_rendering}), we obtain:
\begin{align}\label{ILD_error_analysis_1}
    \epsilon_{ILD}(\Omega,f)  & = \left|20\log{\left(\frac{|h^l s|}{|h^r s|}\frac{|[\mathbf{c}^r]^H  \mathbf{v} s +[\mathbf{c}^r]^H\mathbf{n}|}{|[\mathbf{c}^l]^H  \mathbf{v}s +[\mathbf{c}^l]^H\mathbf{n}|}\right)}\right|  \nonumber\\
               & = \left|20\log{\left(\frac{|h^l|}{|h^r|}\frac{|[\mathbf{c}^r]^H  \mathbf{v}|}{|[\mathbf{c}^l]^H  \mathbf{v}|}\right)}\right| 
\end{align}
The noise component $[\mathbf{c}^{l,r}]^H\mathbf{n}$ in Eq.~\eqref{ILD_error_analysis_1} can be disregarded because ILD analysis typically assumes high SNR.

Next, the BSM coefficients $\mathbf{c}^{l,r}$ are written explicitly in terms of the HRTF, based on the solution structure of Eqs.~\eqref{BSM-MSE},~\eqref{BSM-MLS} where $\mathbf{c}^{l,r}=\mathbf{W}[\mathbf{h}^{l,r}]^*$ and $\mathbf{W} = (\mathbf{V}\mathbf{V}^H + \frac{\sigma_n^2}{\sigma_s^2}\mathbf{I}_M)^{-1} \mathbf{V}$. The method is also referred to as the beamforming-based binaural reproduction (BFBR) approach~\cite{davis2005high}. This substitution leads to
\begin{equation}\label{ILD_error_analysis}
    \epsilon_{ILD}(\Omega,f)  = \left|20\log{\left(\frac{|h^l|}{|h^r|}\frac{|[\mathbf{h}^r]^T\mathbf{W}^H  \mathbf{v}|}{|[\mathbf{h}^l]^T\mathbf{W}^H  \mathbf{v}|}\right)}\right| 
\end{equation}

For $\epsilon_{ILD}(\Omega,f)=0$, the term inside the logarithm must equal $1$, leading to the following equation:
\begin{equation}\label{eq:solvingILD_1}
    |h^l|\, |[\mathbf{h}^r]^T \mathbf{W}^H\mathbf{v}| = |h^r|\, |[\mathbf{h}^l]^T \mathbf{W}^H\mathbf{v}| 
\end{equation}
This can be expressed equivalently as:
\begin{equation}\label{eq:solvingILD_2}
\left\{ \begin{aligned} 
  \left|[\mathbf{h}^r]^T \mathbf{W}^H\mathbf{v}\right| &= \alpha|h^r| \\
  \left|[\mathbf{h}^l]^T \mathbf{W}^H\mathbf{v}\right| &= \alpha|h^l|
\end{aligned} \right.
\end{equation}
where $\alpha \in \mathbb{R}^+$. For the error in Eq.~\eqref{ILD_error_analysis} to be zero for a specific single direction $\Omega$, we assume that the $Q \times 1$ vector $\mathbf{h}^{l,r}$ contains the desired direction $\Omega$ in its $q$-th entry, implying that this entry equals $h^{l,r}$ appearing in the same equation. This assumption is reasonable because the HRTF is provided for a dense set of $Q$ directions.

To design $\mathbf{W}$ that satisfies Eq.~\eqref{eq:solvingILD_2}, the $M \times 1$ columns of $\mathbf{W}$, denoted as $\mathbf{w}_j$, can be selected as follows:
\begin{equation}\label{eq:solvingILD_3}
    \left\{ \begin{aligned} 
  \mathbf{w}_j &\perp \mathbf{v} \quad\text{if} \quad  j \ne q \\
  \mathbf{w}_j &= \frac{\alpha}{\| \mathbf{v} \|^2}\mathbf{v} \quad\text{if} \quad j = q
\end{aligned} \right.
\end{equation}
This choice results in $\mathbf{W}^H\mathbf{v} = [0,\dots,\alpha,\dots,0]^T$, leading to $[\mathbf{h}^{l,r}]^T[0,\dots,\alpha,\dots,0]^T = \alpha h^{l,r}$, thereby satisfying Eq.~\eqref{eq:solvingILD_2}. Designing $\mathbf{W}$ to achieve zero $\epsilon_{ILD}(\Omega,f)$ for a specific $\Omega$ is possible since $\mathbf{v}$ is of size $M \times 1$ while $\mathbf{W}$ is of size $M \times Q$ and so $\mathbf{W}$ has a large null space. 



While designing $\mathbf{W}$ to achieve zero $\epsilon_{ILD}(\Omega,f)$ for a specific $\Omega$ is relatively straightforward, ensuring low ILD error across all $\Omega$ on the horizontal plane is more complex~\cite{xie2013head}. We aim to solve Eq.~\eqref{narrow_band_ILD_e} for a subset $S$ of $Q$ directions on the horizontal plane. Eq.~\eqref{eq:solvingILD_2} can be modified to handle this multi-directional problem:
\begin{equation}\label{eq:solvingILD_4}
\left\{ \begin{aligned} 
  \left|[\mathbf{h}^r]^T \mathbf{W}^H\tilde{\mathbf{V}}\right| &= \alpha|\tilde{\mathbf{h}}^r|^T \\
  \left|[\mathbf{h}^l]^T \mathbf{W}^H\tilde{\mathbf{V}}\right| &= \alpha|\tilde{\mathbf{h}}^l|^T
\end{aligned} \right.
\end{equation}
Here, $\tilde{\mathbf{V}}$ denotes the $M \times Q$ steering matrix, while $\tilde{\mathbf{h}}^{l,r}$ represents $Q \times 1$ vector with $S$ non-zero elements from the horizontal plane. Following the approach for the single direction case in Eq.~\eqref{eq:solvingILD_4}, a solution can be extended to the $S$ sources case. If the null space of $\tilde{\mathbf{V}}$ is non-empty, a zero ILD error $\mathbf{W}$ could be designed; however, when $S \ge M$, the null space of $\tilde{\mathbf{V}}$ may be empty, making a zero ILD error not guaranteed.

From Eq.~\eqref{eq:solvingILD_4}, it is evident that the ILD depends on the magnitudes of $\tilde{\mathbf{h}}^{l,r}$, which represent the listener's HRTF, and $\tilde{\mathbf{V}}$, which encodes the ATF of the microphone array. Additionally, satisfying Eq.~\eqref{eq:solvingILD_4} in practical cases where $M$ is small presents a complex challenge with no closed-form solution. This analysis is sufficient for determining a solution but not necessary, indicating that when $S \gg M$, a solution may not exist, highlighting the difficulty in such cases.


%% file: sections/Ch5_v5.tex
\section{Simulative Study}\label{ch5}
This section presents a detailed performance analysis of the proposed BSM-iMagLS methods and compares it to other state-of-the-art signal independent binaural reproduction methods. It further examines how performance is affected by various factors, such as the microphone array ATF, the HRTF, head-rotation compensation, and reverberation.

\subsection{Setup}\label{ch5:setup}
The signals analyzed in this section are generated through computer simulation. Specifically, they are simulated to represent sound fields consisting of a single plane wave in free field with unit amplitude, with directions of arrival covering the entire directional space for the anechoic condition analysis. Additionally, to evaluate the performance of the methods under reverberant conditions, room impulse responses were simulated using the image method~\cite{allen1979image} and implemented in MATLAB (2023a). Three rectangular rooms were evaluated, with dimensions \([5, 4, 3]\,\text{m},\ [10, 7, 3.5]\,\text{m},\ [20, 15, 6]\,\text{m}\), corresponding to small, medium, and large room sizes, respectively. The rooms reverberation times were \(T_{60} = 0.3, 0.6, 1.4\) seconds, respectively, with a maximum reflection order of $44$. The microphone array was positioned at the center of each room. A single omnidirectional source was randomly placed in each simulation. For each room, $100$ simulations were performed with $100$ different source positions. Based on these room impulse responses, binaural room impulse responses (BRIRs) were generated for each method. The analysis is performed for several HRTF datasets, including the Cologne HRTF database for the Neumann KU100 dummy head~\cite{bernschutz2013spherical} and the HUTUBS database, which features data from 96 subjects~\cite{ brinkmann2019hutubs}. 

The signals are evaluated using four microphone arrays, as outlined in Table~\ref{tab:Mic_arrays}. The ATF of the circular and semi-circular arrays was simulated using an analytic model of a rigid sphere~\cite{rafaely2015fundamentals}, with microphones arranged on the sphere's horizontal plane at equal angular intervals. The twelve-microphone array represents an idealized configuration covering the entire azimuth space, while the six-microphone array approximates a head-mounted array. The ATFs of the other two, which are arrays mounted on glasses, were measured on a dummy head in an anechoic chamber, simulating a realistic head-mounted microphone setup. The four-microphone glasses array is part of the publicly available EasyCom dataset~\cite{donley2021easycom}.

The ATF for the five-microphone glasses configuration was measured at $1625$ directions, using a grid approximately conforming to a Lebedev grid of order $35$. The four-microphone EasyCom ATF was measured at $1020$ directions, following a Lebedev grid of order $29$. Both the circular and semi-circular arrays, with a radius of $10$ cm, were simulated with ATFs having $1625$ directions, similar to the five-microphone head-mounted array.

\begin{table}[ht!]
    \caption{Summary of microphone arrays used in the evaluation.}
    \label{tab:Mic_arrays}
    \centering
    \begin{tabular}{|p{2.0cm}|p{1.5cm}|p{1.5cm}|p{1.6cm}|}
        \hline
        Array Description&Number of Microphones&Simulated / Measured&Mounted on\\
        \hline
        \hline
        Circular&12&Simulated&Rigid sphere\\
        \hline
        Semi-circular&6&Simulated&Rigid sphere\\
        \hline
        Glasses&5&Measured&Dummy head\\
        \hline
        EasyCom&4&Measured&Dummy head\\
        \hline
    \end{tabular}

\end{table}

\subsection{Methodology}\label{sec:5_Methodology}
This section provides a detailed description of the different binaural signals used in the analysis.

\subsubsection{$p^{l,r}$} Reference left and right binaural signals, calculated as per Eq.~\eqref{true_binaural_signals}. For a single plane wave in free field, the reference signals are equivalent to the measured or simulated HRTF.

\subsubsection{\(z^{l,r}_{LS}\)} BSM-LS binaural signals were computed using Eq.~\eqref{BSM-MSE} and Eq.~\eqref{BSM_rendering}, with an FFT size of $1024$. The ATF matrix $\mathbf{V}$ and the HRTF vector $\mathbf{h}^{l,r}$ were defined based on the evaluated array configuration and HRTF. An SNR of \(20\,\text{dB}\) was used for this evaluation.

\subsubsection{\(z^{l,r}_{MLS}\)} BSM-MagLS binaural signals, computed using Eq.~\eqref{BSM-MLS} and Eq.~\eqref{BSM_rendering} and the same FFT size of BSM-LS. The BSM-MagLS coefficients in Eq.~\eqref{BSM-MLS} are determined using $f_c = 1.5$ kHz, and the ATF matrix $\mathbf{V}$ is determined according to the evaluated array configuration. The SNR used in this evaluation is set to $20$ dB.

\subsubsection{\(z^{l,r}_{eMLS}\)} The eMagLS binaural signals—an end-to-end variant of MagLS proposed in~\cite{deppisch2021end}—were computed following the implementation in~\cite{deppisch2024blind}, using the same FFT size, SNR, and cutoff frequency as the BSM-MagLS signals.

\subsubsection{$z^{l,r}_{iMLS}$} 
\begin{figure}
    \centering
    \includegraphics[width=0.45\textwidth,trim={3cm 0.5cm 4cm 0.5cm},clip]{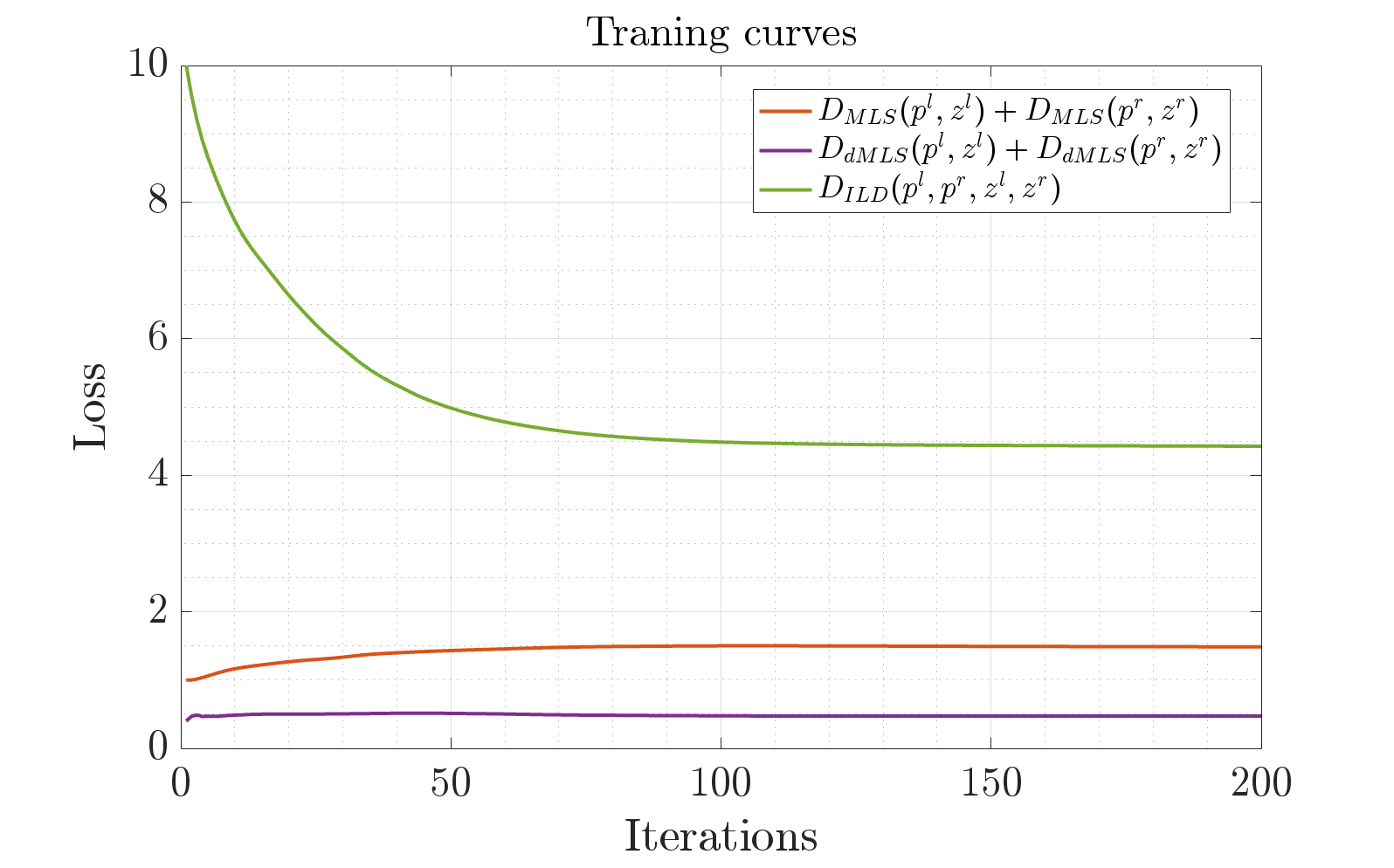}
    \caption{The three BSM-iMagLS error components of Eq.~\eqref{D_iMLS} as a function of training iteration for KU100 HRTF and the EasyCom ATF. In this example the weighting vector is set to $\mathbf{\lambda}=[0.4,10]$.}
    \label{fig:learning_curves}
\end{figure}

BSM-iMagLS binaural signals, generated using the proposed method described in Section~\ref{sec:3}. Here, BSM-MagLS coefficients are used as the initial solution for the DNN. The weighting factors $\mathbf{\lambda}_{1,2}$ for derivative magnitude and ILD errors are set to $[0.4, 10]$, respectively. $\mathbf{\lambda}_{1,2}$ were chosen empirically based on preliminary experiments to balance the influence of the error factors on the overall loss. The learning rate is set to $0.0008$, with $200$ iterations allocated for error convergence, although in this case, convergence was achieved in approximately $100$ iterations. The BSM-iMagLS coefficient computation takes approximately three minutes on an Apple M2 Max MacBook Pro using PyTorch. Figure~\ref{fig:learning_curves} shows the learning curves of the weighted errors over iterations, illustrating a significant decrease in ILD error while other errors remain relatively stable. One possible reason for the smoothness of the training curves is that the optimization is performed over a single HRTF/ATF which reduces the stochasticity typically present in batch-based training. The relatively stable behavior of the $D_\text{MLS}$ and $D_\text{dMLS}$ errors, compared to the ILD error, can likely be attributed to the fact that the initial BSM coefficients are already set to the optimized BSM-MagLS solution. Starting from a favorable point on the optimization surface can accelerate overall convergence, especially for errors that are already near their minima.

\subsection{Evaluation Metrics}\label{sec5:Metrics}
The performance of methods is evaluated by comparing the signals $z^{l,r}_{MLS}$, $z^{l,r}_{eMLS}$,  and $z^{l,r}_{iMLS}$ to the reference $p^{l,r}$ using five objective metrics:

\subsubsection{Normalized Mean Square Error (NMSE)} is defined as:
\begin{equation}\label{eq:NMSE-anal}
    \epsilon^{l,r}_{nmse}(f,\Omega) = 10\log_{10}\frac{|p^{l,r}(f,\Omega) - z^{l,r}(f,\Omega)|^2}{|p^{l,r}(f,\Omega)|^2},
\end{equation}
where $z^{l,r}$ denotes the evaluated signal, either BSM-MagLS or BSM-iMagLS. The NMSE is evaluated for $2702$ directions on a dense Lebedev grid of order $44$. This metric captures both magnitude and phase errors, making it a reliable measure of the binaural signal's physical accuracy.

\subsubsection{Normalized Magnitude Error} is defined as:
\begin{equation}\label{eq:mag-anal}
    \epsilon_{mag}^{l,r}(f,\Omega) = 10\log_{10}\frac{|\,\,|p^{l,r}(f,\Omega)| - |z^{l,r}(f,\Omega)|\,\,|^2}{|p^{l,r}(f,\Omega)|^2}.
\end{equation}
This measure emphasizes the importance of magnitude accuracy, and is particularly relevant at high frequencies, which is crucial for perceptual sound localization \cite{wightman1992dominant}.

\subsubsection{ILD Error} denoted $\epsilon_{ILD}(\Omega)$ is computed using Eq.~\eqref{ILD_error_eq} and is evaluated over $361$ equally spaced directions on the horizontal plane. The ILD is assessed from $1.5$ kHz to $20$ kHz across $23$ auditory filter bands. 

\subsubsection{BSD and LSD} 
The basic spectral difference (BSD) is defined as:
\begin{equation}
    \epsilon_{BSD}^{l,r}(f_{i}) = 10 \log_{10}\frac{|ERB\left(z^{l,r},f_{i}\right)|^2}{|ERB\left(p^{l,r},f_{i}\right)|^2}
\end{equation}
and is evaluated over $23$ auditory filters with centered frequencies $f_{i}$ distributed on an equivalent
rectangular bandwidth (ERB) frequency scale between $1.5$ kHz and $20$ kHz. The log spectral distance (LSD) is defined as:
\begin{equation}
    \epsilon_{LSD}^{l,r} =\sqrt{\frac{1}{23}\sum_{f_{i}=1}^{23} \left(\epsilon_{BSD}^{l,r}(f_{i}) \right)^2 }
\end{equation}
BSD and LSD operate on level differences in decibels of each frequency band on the filtered BRIRs as an approximation for coloration differences~\cite{mckenzie2025toward}.

\subsection{Results}

The results are presented in four parts, each focusing on the comparison of BSM-LS, BSM-MagLS, eMagLS, and BSM-iMagLS under different conditions, including various ATFs, HRTFs, and head-rotation compensation.

\subsubsection{Performance for the Four Microphone Arrays}\label{ch5a}
As discussed in Section~\ref{sec:4}, the mitigation of ILD error depends on the microphone array ATF matrix $\mathbf{V}$. The three methods described in Sec.~\ref{sec:5_Methodology} were used to design filters for the four arrays described in Sec.~\ref{ch5:setup} and the KU100 HRTF. Performance was evaluated in terms of NMSE, magnitude error, and ILD error, as described in Sec.\ref{sec5:Metrics}.

\begin{figure}
    \centering
    \subfigure[12 mic circular]{\includegraphics[width=0.425\textwidth,trim={2cm 2cm 2cm 1cm},clip]{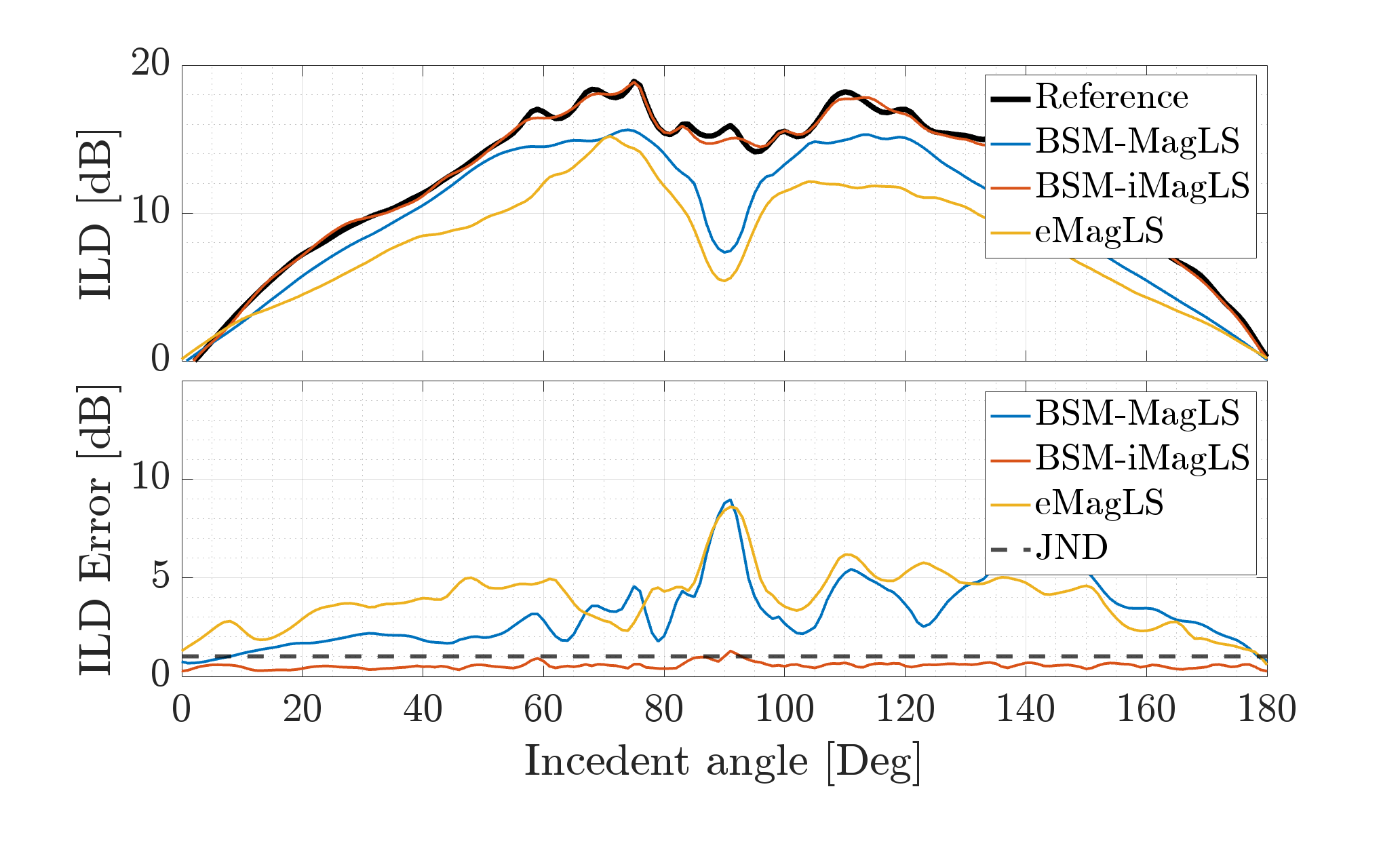}\label{fig:array_config_ILD_ang_12}}
    \subfigure[6 mic semi-circular]{\includegraphics[width=0.425\textwidth,trim={2cm 2cm 2cm 1cm},clip]{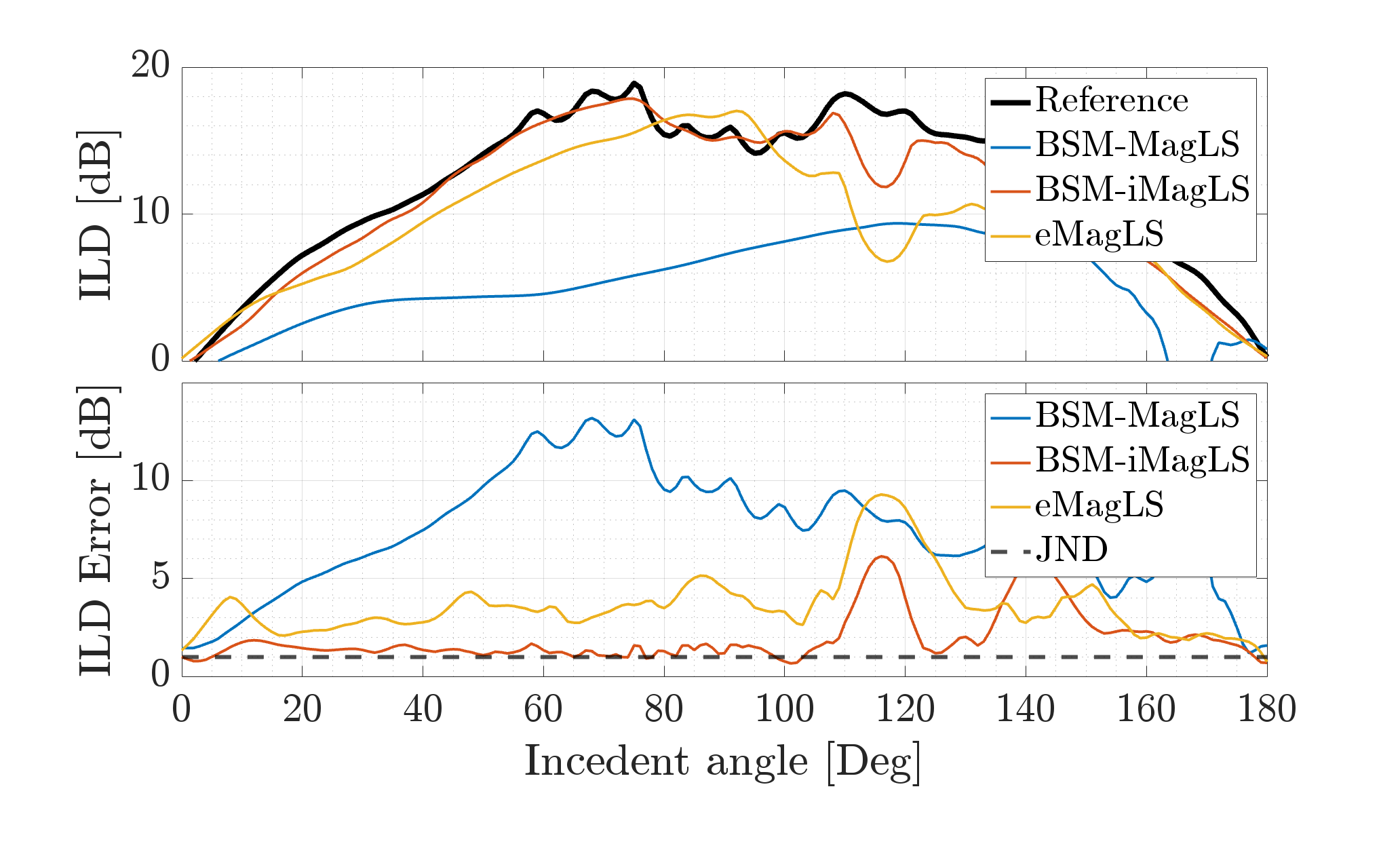}\label{fig:array_config_ILD_ang_6}}
    \subfigure[5 mic wearable glasses array]{\includegraphics[width=0.425\textwidth,trim={2cm 2cm 2cm 1cm},clip]{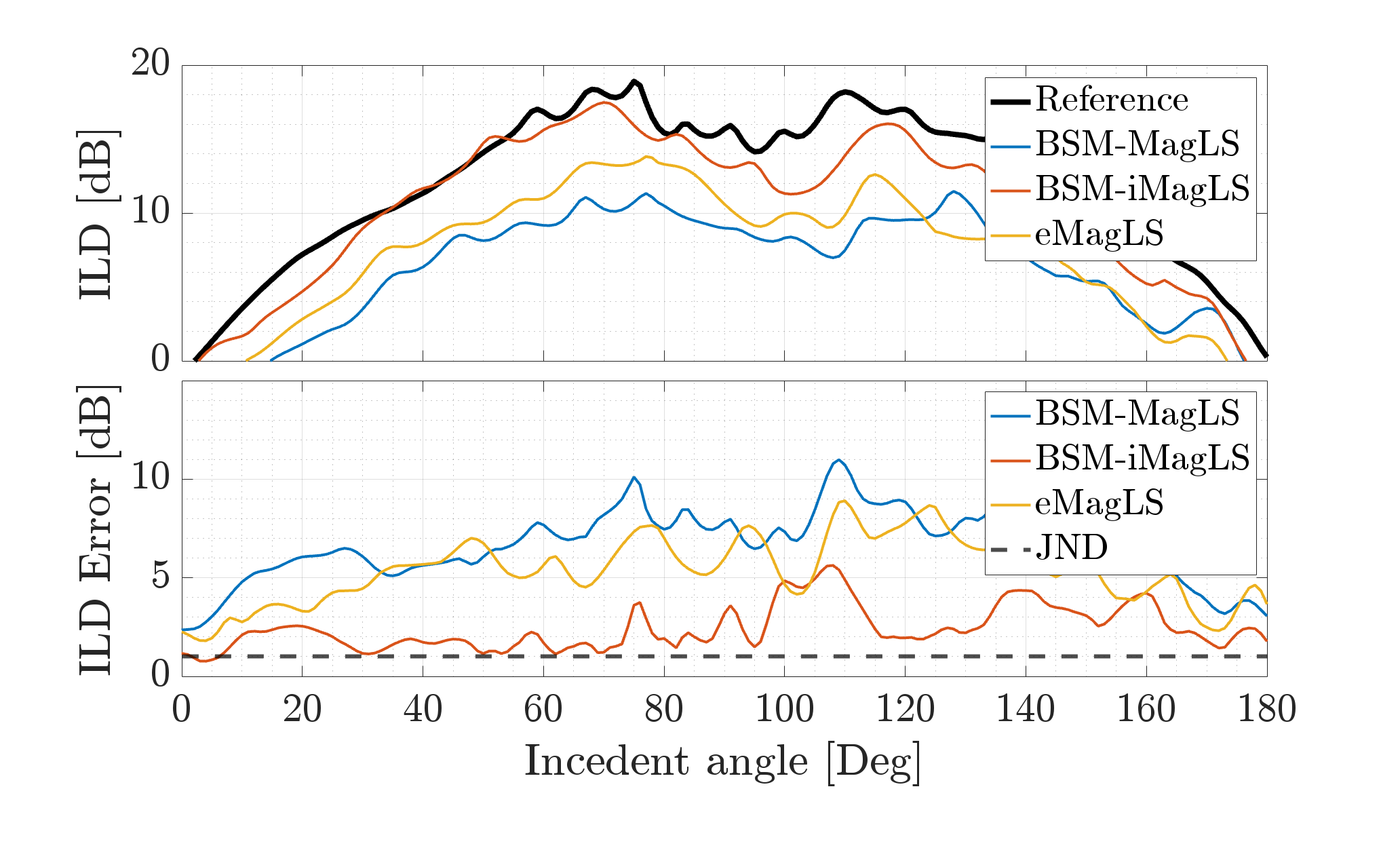}\label{fig:array_config_ILD_ang_5}}
    \subfigure[4 mic EasyCom array]{\includegraphics[width=0.425\textwidth,trim={2cm 2cm 2cm 1cm},clip]{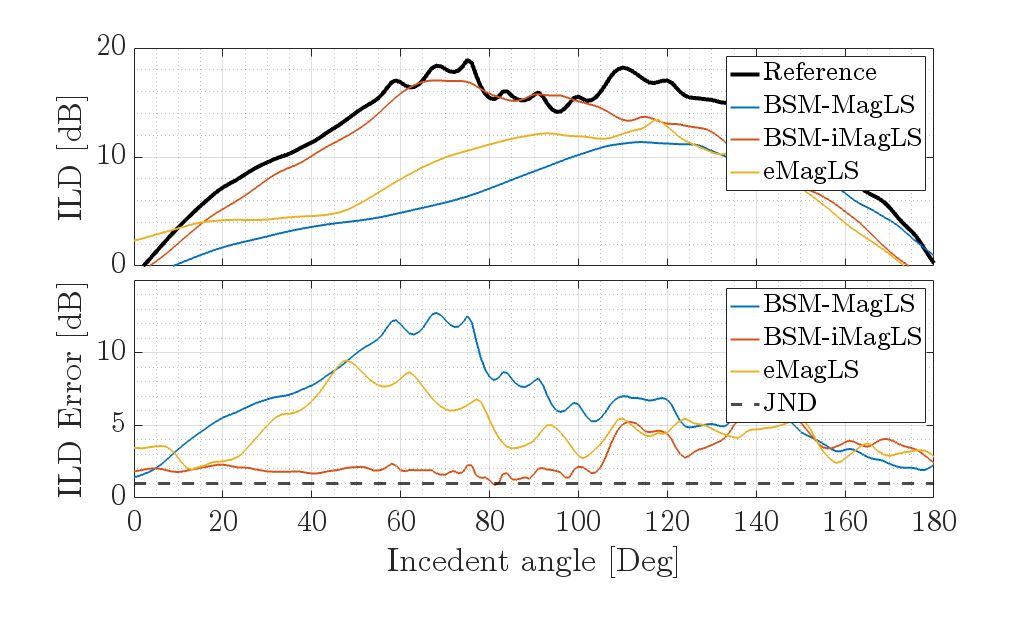}\label{fig:array_config_ILD_ang_4}}
    \caption{ILD curves (top) and ILD error (bottom) as a function of incident angle for the four microphone array configurations.}
    \label{fig:array_config_ILD_ang}
\end{figure}

Figure~\ref{fig:array_config_ILD_ang}  presents the ILD curves (top) and errors (bottom) as a function of incident angle (\(0^\circ-180^\circ\)) for each rendering method and array configuration. Across all arrays, BSM-iMagLS consistently reduces ILD error compared to the other methods, with the most significant improvements observed in frontal and contralateral directions, where microphones are present. BSM-iMagLS manage to lower the ILD error close to the Just Noticeable Difference (JND) threshold of \(1\) dB~\cite{yost1988discrimination} in most cases and angles, highlighting its robustness across varying configurations.

\begin{figure}
    \centering
    \subfigure[12 mic circular]{\includegraphics[width=0.425\textwidth,trim={1.7cm 2cm 2cm 1cm},clip]{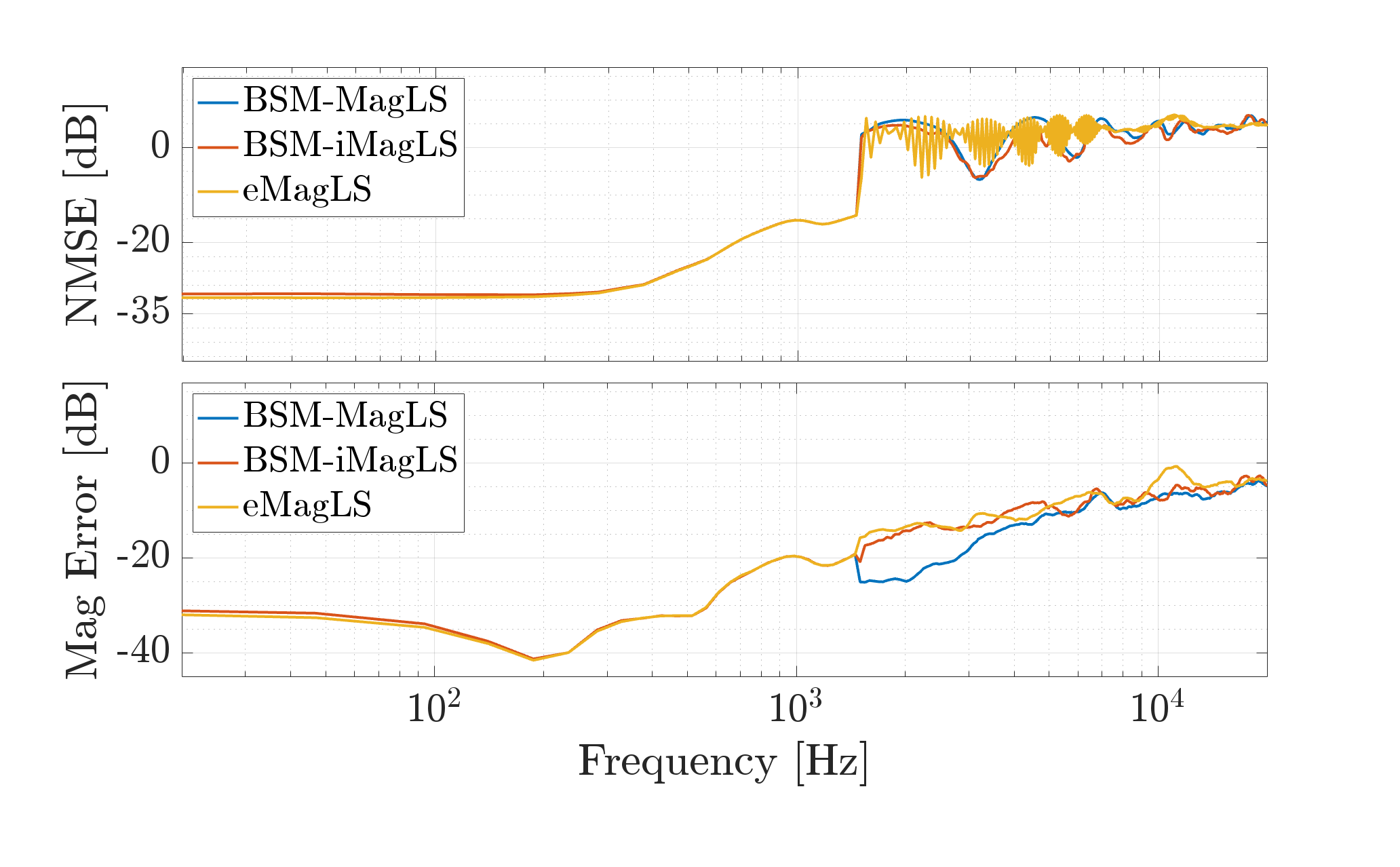}\label{fig:array_config_nmse_mag_ang_12}}
    \subfigure[6 mic Semi circular]{\includegraphics[width=0.425\textwidth,trim={1.7cm 2cm 2cm 1cm},clip]{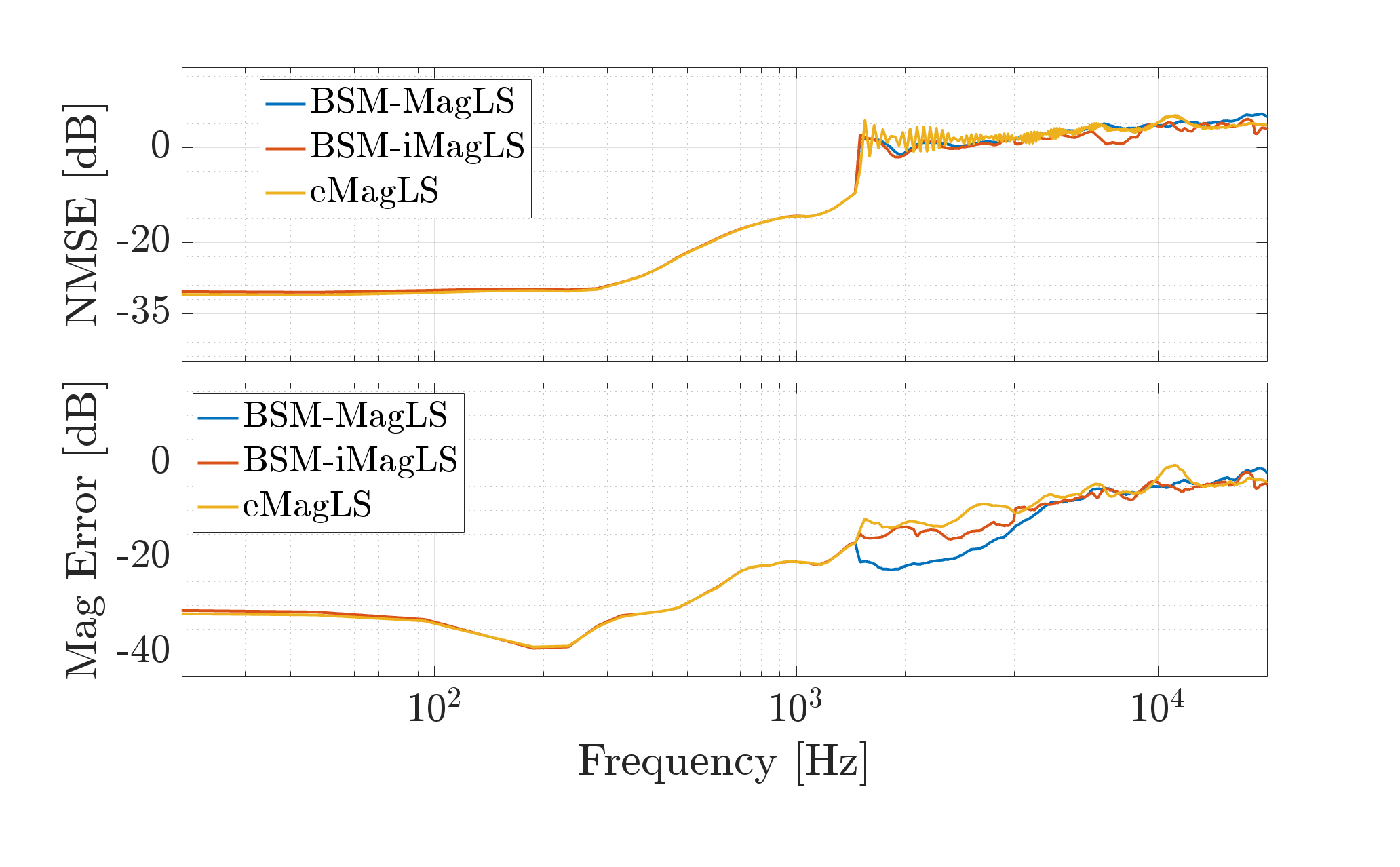}\label{fig:array_config_nmse_mag_ang_6}}
    \subfigure[5 mic wearable glasses array]{\includegraphics[width=0.425\textwidth,trim={1.7cm 2cm 2cm 1cm},clip]{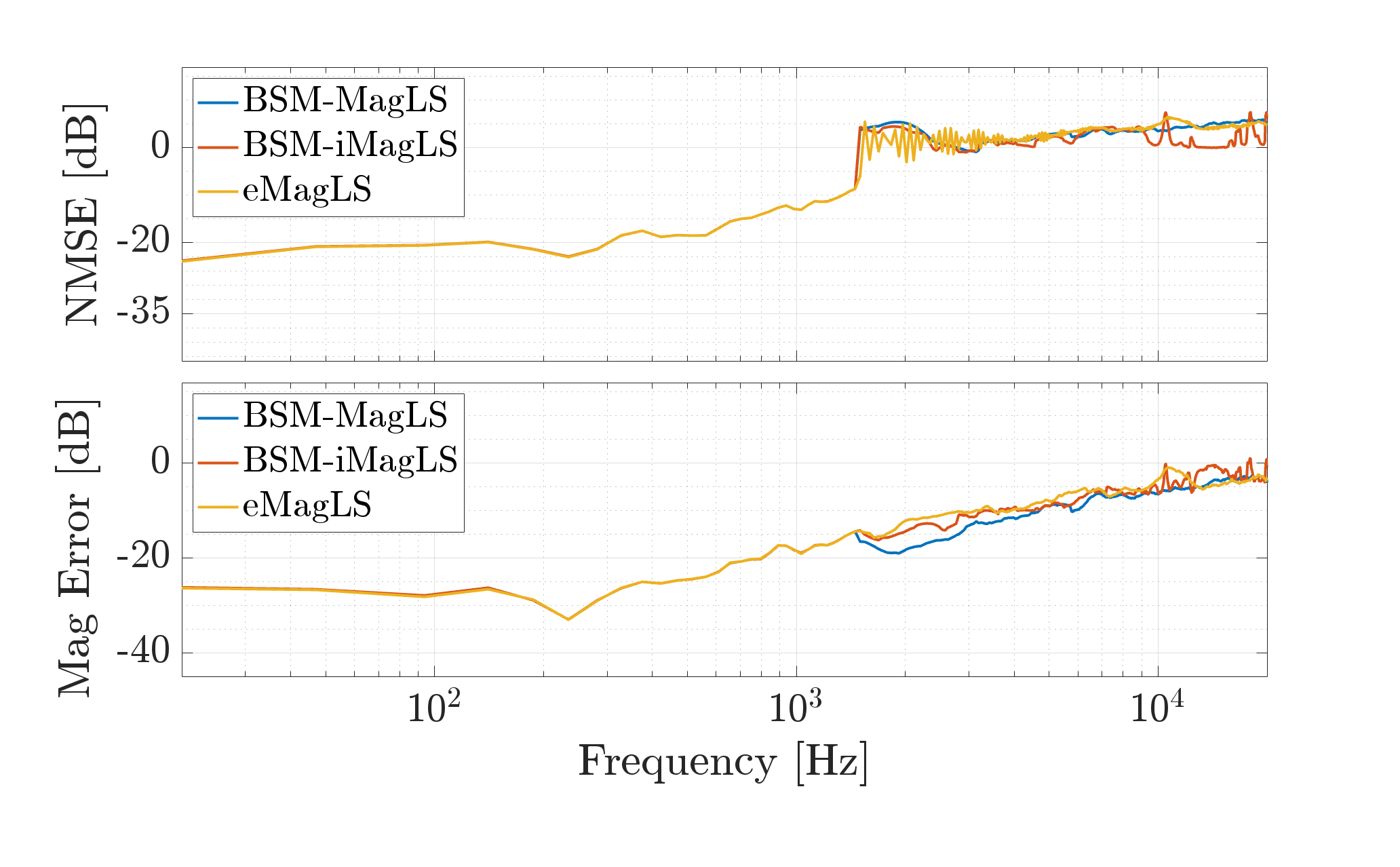}\label{fig:array_config_nmse_mag_ang_5}}
    \subfigure[4 mic EasyCom array]{\includegraphics[width=0.425\textwidth,trim={1.7cm 2cm 2cm 1cm},clip]{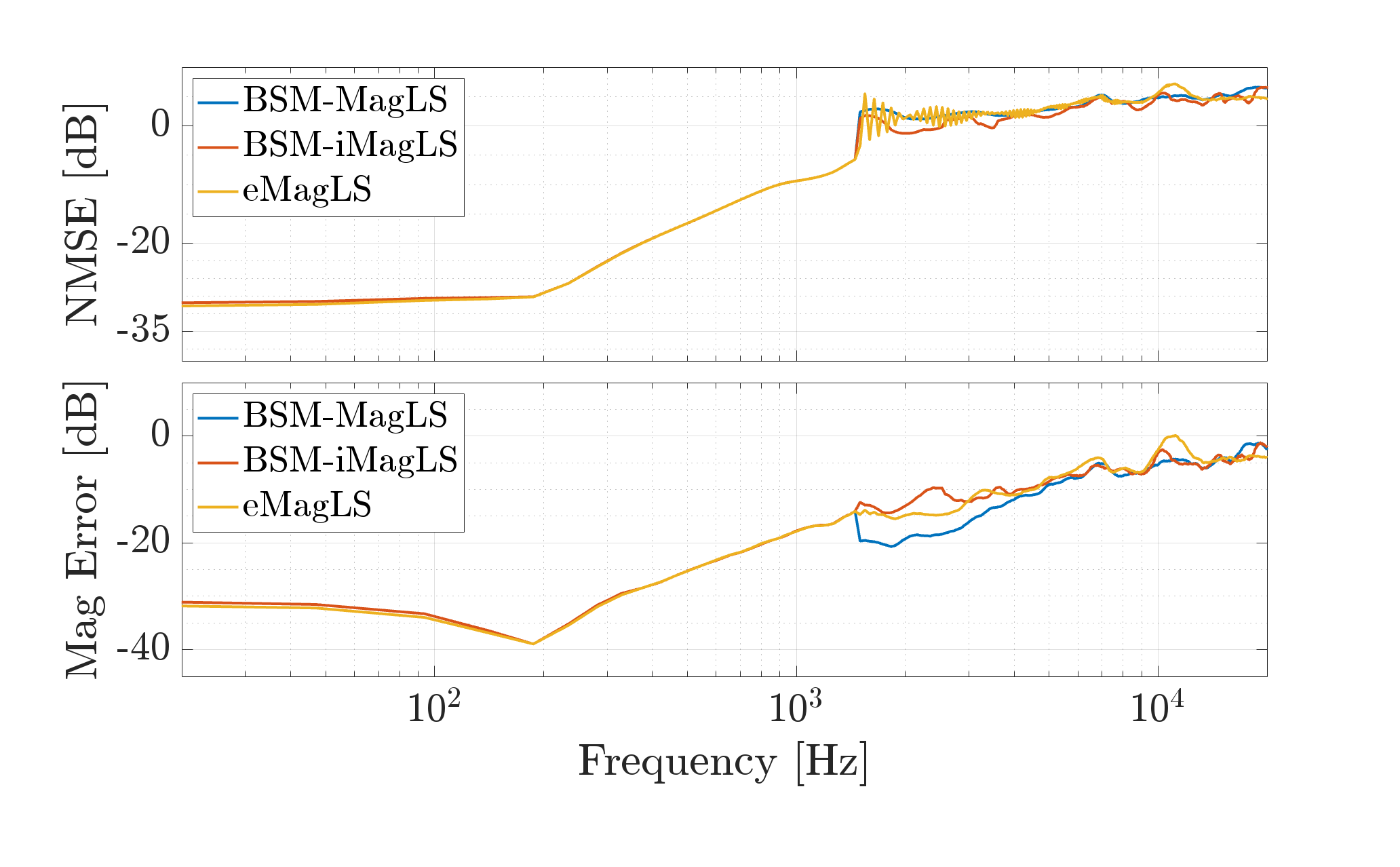}\label{fig:array_config_nmse_mag_ang_4}}
    \caption{NMSE (top) and Normalized Magnitude error (bottom) as a function of frequency for four microphone array configurations.}
    \label{fig:array_config_nmse_mag}
\end{figure}

Next, Fig.~\ref{fig:array_config_nmse_mag} presents the NMSE (top) and magnitude error (bottom) for the evaluated methods across all array configurations. The NMSE below $1.5$kHz is identical for all methods as expected from Sec.~\ref{sec:3}, while above that frequency the error is high for the methods as they do not preserve phase (Eqs.~\eqref{eq:mls_phase},~\eqref{D_iMLS}). Magnitude error, more perceptually relevant above \(1.5\) kHz, is slightly higher for BSM-iMagLS than BSM-MagLS across all configurations and comparable to eMagLS error. This increase is more pronounced in the \(1.5-4\) kHz range, but error levels still remain below \(-10\) dB. For the remainder of the frequency range, trends are similar, with the exception of the five and four-microphone arrays, which show increased noise above \(10\) kHz, likely due to measurement noise at those high frequencies.

These results highlight that performance is dependent on the array configuration. Across all cases, BSM-iMagLS significantly reduces ILD errors compared to the other methods, though with a small increase in magnitude error. This trade-off could be advantageous in terms of perception, particularly in azimuth source localization ~\cite{xie2013head,aronoff2010use}.

\begin{table}
    \caption{Summary of the $\text{mean}_{(\text{STD})}$ errors across the HUTUBS dataset for four microphone array configurations.}
    \label{tab:metrics_summary}
    \centering
    \resizebox{\columnwidth}{!}{
    \begin{tabular}{|>{\centering\arraybackslash}m{2.7cm}|c|c|c|c|}
        \hline
        \textbf{Array Configuration} & \textbf{Method} & \textbf{ILD [dB]} & \textbf{Mag [dB]} & \textbf{NMSE [dB]} \\
        \hline
        \multirow{2}{*}{\textbf{12 Mic Circular}} & BSM-MagLS & $3.84_{(0.86)}$ & $-16.02_{(1.25)}$ & $-3.39_{(0.86)}$ \\
                                                  & BSM-iMagLS & $0.74_{(0.18)}$ & $-14.05_{(1.47)}$ & $-4.08_{(1.18)}$ \\
        \hline
        \multirow{2}{*}{\textbf{6 Mic Semi-Circular}} & BSM-MagLS & $6.93_{(1.24)}$ & $-13.94_{(1.15)}$ & $-2.57_{(0.59)}$ \\
                                                      & BSM-iMagLS & $2.26_{(0.58)}$ & $-13.31_{(1.20)}$ & $-3.28_{(0.72)}$ \\
        \hline
        \multirow{2}{*}{\textbf{5 Mic Glasses}} & BSM-MagLS & $6.32_{(0.98)}$ & $-13.81_{(0.85)}$ & $-3.19_{(0.47)}$ \\
                                                & BSM-iMagLS & $2.32_{(0.47)}$ & $-12.39_{(1.08)}$ & $-4.35_{(0.77)}$ \\
        \hline
        \multirow{2}{*}{\textbf{4 Mic EasyCom}} & BSM-MagLS & $6.45_{(1.43)}$ & $-13.66_{(1.12)}$ & $-2.68_{(0.58)}$ \\
                                                & BSM-iMagLS & $3.10_{(0.75)}$ & $-12.62_{(1.25)}$ & $-2.73_{(0.77)}$ \\
        \hline
    \end{tabular}
    }
\end{table}

\subsubsection{Performance Over a Large HRTF Dataset}
This section examines the influence of various HRTFs on the proposed method. Specifically, performance is evaluated using the HUTUBS database, and the ATFs of Table.~\ref{tab:Mic_arrays}, as detailed in Sec.~\ref{ch5:setup}.

\begin{figure}
    \centering
    \subfigure[ILD error as a function of incident angle]{\includegraphics[width=0.45\textwidth,trim={0cm 0cm 0cm 0cm},clip]{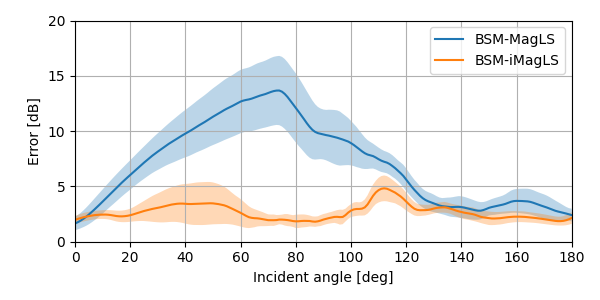}\label{fig:HRTF_analysis_ILD_ang}}
    \subfigure[ILD error as a function of frequency]{\includegraphics[width=0.45\textwidth,trim={0cm 0cm 0cm 0cm},clip]{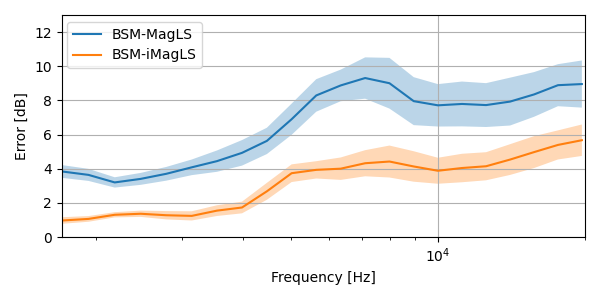}\label{fig:HRTF_analysis_ILD_freq}}
    \subfigure[NMSE]{\includegraphics[width=0.45\textwidth,trim={0cm 0cm 0cm 0cm},clip]{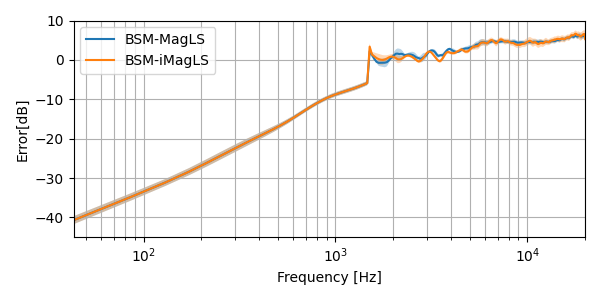}\label{fig:HRTF_analysis_nmse_freq}}
    \subfigure[Normalized magnitude error]{\includegraphics[width=0.45\textwidth,trim={0cm 0cm 0cm 0cm},clip]{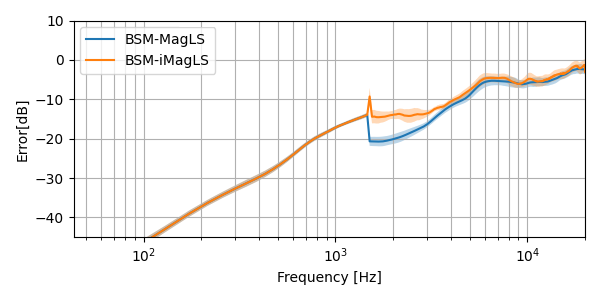}\label{fig:HRTF_analysis_mag_freq}}
    \caption{Error analysis for the HUTUBS HRTF data set and the EasyCom array ATF showing (a) ILD error, as a function of direction (b) ILD error as a function of frequency (c) NMSE as a function of frequency and (d) Normalized magnitude error as a function of frequency. The curves depicts mean and standard deviation with regards to the subjects in the HUTUBS dataset.}
    \label{fig:HRTF_analysis}
\end{figure}

Figure~\ref{fig:HRTF_analysis_ILD_ang} shows the ILD error as a function of incident angle for the EasyCom ATF. The lines indicates the mean and standard deviation over all HRTFs in the database. The BSM-iMagLS curve consistently demonstrates lower ILD errors, particularly for frontal angles, with a low standard deviation. This indicates that the proposed method offers consistent performance across a diverse range of HRTFs. Importantly, an overall improvement is observed across all HRTFs, as the curves largely do not intersect for most incident angles. This trend is further confirmed in Figure~\ref{fig:HRTF_analysis_ILD_freq}, where the ILD error is shown as a function of frequency. A clear reduction in ILD error is observed across all frequencies and HRTFs, supporting the consistency of BSM-iMagLS in improving ILD error.

Figures~\ref{fig:HRTF_analysis_nmse_freq} and~\ref{fig:HRTF_analysis_mag_freq} present the NMSE and magnitude errors as functions of frequency, for the EasyCom ATF. The NMSE remains relatively stable across all frequencies, while the magnitude error decreases to below \(-10\) dB within the \(1.5 - 3\) kHz range. Outside this range, the magnitude error remains smooth and consistent for both BSM-MagLS and BSM-iMagLS. These results align with previous findings, indicating minimal impact on both magnitude and complex errors.

A summary of the NMSE, magnitude, and ILD errors is provided in Table~\ref{tab:metrics_summary} for the other array configurations. The mean and standard deviations across the HRTFs in the set, are presented. Notably, NMSE and magnitude errors were assessed only for frequencies above \(1.5\) kHz, as both BSM-MagLS and BSM-iMagLS are identical below this threshold.

Results in Table~\ref{tab:metrics_summary} for ILD error, BSM-MagLS remains around \(6-7\) dB across all array configurations. In contrast, BSM-iMagLS achieves errors of around \(2-3\) dB with lower standard deviations. The NMSE values for both methods are found to be similar, with the magnitude error increasing by approximately \(1\) dB across all arrays. These results further support that BSM-iMagLS is effective across a broad spectrum of HRTFs.

\begin{table}[ht!]
    \caption{Summary of the $\text{mean}_{(\text{STD})}$ errors with head rotation compensation of the KU100 HRTF for four microphone array configurations.}
    \label{tab:metrics_summary_head_traking}
    \centering
    \resizebox{\columnwidth}{!}{
    \begin{tabular}{|>{\centering\arraybackslash}m{2.7cm}|c|c|c|}
        \hline
        \textbf{Array Configuration} & \textbf{Method} & \textbf{ILD [dB]} & \textbf{Mag [dB]} \\
        \hline
        \multirow{3}{*}{\textbf{12 Mic Circular}} & BSM-MagLS & $3.52_{(0.17)}$ & $-16.77_{(0.19)}$ \\
                                                  & BSM-iMagLS & $0.64_{(0.04)}$ & $-15.24_{(0.08)}$ \\
                                                  & eMagLS & $4.81_{(0.24)}$ & $-13.42_{(0.05)}$ \\
        \hline
        \multirow{3}{*}{\textbf{6 Mic Semi-Circular}} & BSM-MagLS & $6.81_{(0.61)}$ & $-14.67_{(0.43)}$ \\
                                                      & BSM-iMagLS & $3.21_{(0.86)}$ & $-14.03_{(0.54)}$ \\
                                                      & eMagLS & $8.91_{(1.51)}$ & $-10.99_{(0.62)}$ \\
        \hline
        \multirow{3}{*}{\textbf{5 Mic Glasses}} & BSM-MagLS & $7.37_{(1.03)}$ & $-13.95_{(0.44)}$ \\
                                                & BSM-iMagLS & $2.99_{(0.79)}$ & $-12.54_{(0.59)}$ \\
                                                & eMagLS & $9.11_{(1.80)}$ & $-11.28_{(0.61)}$ \\
        \hline
        \multirow{3}{*}{\textbf{4 Mic EasyCom}} & BSM-MagLS & $7.06_{(0.86)}$ & $-14.05_{(0.48)}$ \\
                                                & BSM-iMagLS & $3.63_{(0.78)}$ & $-13.06_{(0.59)}$ \\
                                                & eMagLS & $7.99_{(1.74)}$ & $-11.64_{(0.65)}$ \\
        \hline
    \end{tabular}
    }
\end{table}

\subsubsection{Performance Under Head Rotation Compensation}\label{ch5_head_tracking}
This section evaluates the performance of the proposed method under head rotation compensation. In this case the listener head movements during playback are counteracted by aligning the HRTF orientation with the listener’s head using spherical harmonics rotation~\cite{rafaely2008spherical}. This ensures stable source localization in immersive playback, as discussed in~\cite{berebi2023analysis, berebi2021enabling, madmoni2025design}. Implementing this head-tracked BSM scheme requires input from a head tracking device, along with the pre-calculation of BSM coefficients for each head orientation. Afterward, the appropriate BSM coefficients can be selected for Eq.~\eqref{BSM_rendering} based on the data provided by the head tracker.

Despite the feasibility of head tracking compensation, certain challenges arise when applying it to BSM rendering. The primary challenge is that when the HRTF is rotated, the virtual ear positions shift may lead to a new minimization problem that can be more difficult to solve. This issue is particularly pronounced when the ears of the listener are positioned farther from the microphones, which can negatively affect ILD performance~\cite{madmoni2025design}.

\begin{figure}
    \centering
    \includegraphics[width=0.45\textwidth,trim={1cm 2cm 2.5cm 1.3cm},clip]{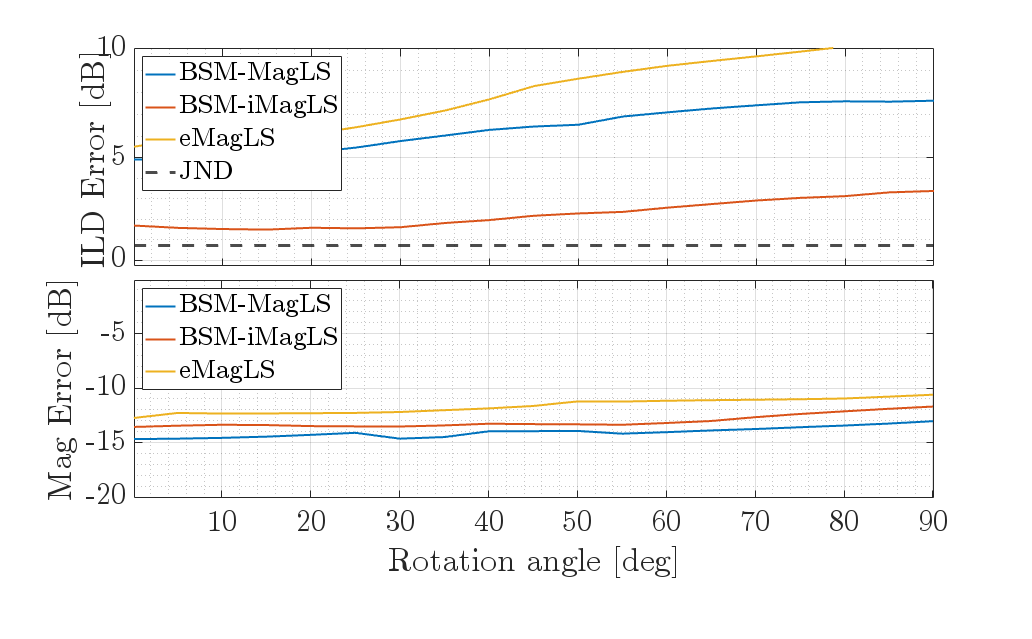}
    \caption{Head rotation compensation analysis, performed on KU100 and the measured EasyCom array. Top figure shows the averaged ILD error for each rotation angle from $0^\circ-90^\circ$, bottom figure shows averaged magnitude error for each rotation angle.} 
    \label{fig:head_rotation_analysis}
\end{figure}

To clarify the evaluation setup in Figure~\ref{fig:head_rotation_analysis}, the reported ILD (top) and magnitude errors (bottom) are computed with respect to a reference HRTF that has been counter-rotated to match the listener’s head orientation. Specifically, as the head rotates on the horizontal plane, the HRTF is rotated accordingly using spherical harmonics rotation to simulate a head-tracked playback scenario, where the auditory scene remains perceptually stable, or world-locked. Meanwhile, the microphone array remains fixed in its initial orientation. This configuration isolates the effect of head movement and allows assessment of how well the BSM method maintains accurate spatial cues under head-rotation compensation. The resulting errors therefore reflect the mismatch between the BSM output and the expected target after head-tracked compensation has been applied.

Figure~\ref{fig:head_rotation_analysis} presents the ILD and magnitude errors averaged over frequency and direction, for head rotations between \(0^\circ\) and \(90^\circ\), using the EasyCom head-mounted array and the KU100 HRTF. BSM-iMagLS consistently outperforms BSM-MagLS and eMagLS in ILD error, achieving a reduction of approximately \(4\) dB across all rotation angles. Notably, ILD error for BSM-iMagLS remains near the JND threshold at smaller rotation angles, demonstrating its robustness in maintaining accurate lateralization.

Magnitude error within the \(1.5-20\) kHz range, shown in the lower figure, reveals a slight advantage for BSM-MagLS, with a consistent \(1\) dB lower error compared to BSM-iMagLS while eMagLS results in the highest averaged magnitude error per head compensation angle. This minor trade-off aligns with trends observed in previous sections, where ILD improvements were prioritized over magnitude fidelity.

Table~\ref{tab:metrics_summary_head_traking} summarizes the mean ILD and magnitude errors for all array configurations under head rotation. Across all cases, BSM-iMagLS significantly reduces ILD errors while incurring a modest increase in magnitude error. These findings emphasize the potential of BSM-iMagLS to enhance localization accuracy under head-tracked conditions, offering a perceptual advantage in dynamic playback scenarios.

\begin{table}
    \caption{Summary of the $\text{mean}_{(\text{STD})}$ LSD error with reverberant BRIRs compensation of the KU100 HRTF for four microphone array configurations.}
    \label{tab:metrics_summary_BRIRs}
    \centering
    \resizebox{\columnwidth}{!}{
    \begin{tabular}{|>{\centering\arraybackslash}m{1.7cm}|
                >{\centering\arraybackslash}m{1.7cm}|
                >{\centering\arraybackslash}m{1.7cm}|
                >{\centering\arraybackslash}m{1.7cm}|
                >{\centering\arraybackslash}m{1.7cm}|}
        \hline
        \shortstack{\textbf{Array} \\ \textbf{Configuration}} & \textbf{Method} & \shortstack{\textbf{Small Room} \\ \textbf{LSD [dB]}} & \shortstack{\textbf{Medium Room} \\ \textbf{LSD [dB]}} & \shortstack{\textbf{Large Room} \\ \textbf{LSD [dB]}} \\
        \hline
        \multirow{4}{*}{\shortstack{\textbf{12 Mic} \\ \textbf{Circular}}} & BSM-LS & $8.06_{(0.45)}$ & $7.87_{(0.47)}$ & $8.03_{(0.81)}$  \\
                                                  & BSM-MagLS & $3.76_{(0.51)}$ & $3.68_{(0.46)}$ & $3.83_{(0.68)}$ \\
                                                  & BSM-iMagLS & $4.14_{(0.56)}$ & $4.07_{(0.49)}$ & $4.21_{(0.77)}$  \\
                                                  & eMagLS & $5.76_{(0.64)}$ & $5.64_{(0.7)}$ & $5.82_{(0.9)}$  \\
        \hline
        \multirow{4}{*}{\shortstack{\textbf{6 Mic} \\ \textbf{Semi-Circular}}} & BSM-LS & $19.32_{(0.41)}$ & $19.08_{(0.62)}$ & $19.09_{(0.88)}$ \\
                                                      & BSM-MagLS & $3.82_{(0.3)}$ & $3.71_{(0.37)}$ & $3.81_{(0.43)}$  \\
                                                      & BSM-iMagLS & $5.22_{(0.5)}$ & $5.01_{(0.64)}$ & $5.15_{(0.75)}$  \\
                                                      & eMagLS & $4.95_{(0.52)}$ & $4.74_{(0.67)}$ & $4.91_{(0.78)}$ \\
        \hline
        \multirow{4}{*}{\shortstack{\textbf{5 Mic} \\ \textbf{Glasses}}} & BSM-LS & $15.94_{(0.82)}$ & $15.94_{(0.92)}$ & $15.92_{(1.18)}$ \\
                                                & BSM-MagLS & $3.56_{(0.55)}$ & $3.63_{(0.62)}$ & $3.66_{(0.74)}$ \\
                                                & BSM-iMagLS & $6.75_{(0.68)}$ & $6.79_{(0.72)}$ & $6.88_{(0.87)}$  \\
                                                & eMagLS & $6.21_{(0.73)}$ & $6.32_{(0.92)}$ & $6.34_{(1.07)}$  \\
        \hline
        \multirow{4}{*}{\shortstack{\textbf{4 Mic} \\ \textbf{EasyCom}}} & BSM-LS & $20.37_{(0.79)}$ & $20.46_{(0.8)}$ & $19.83_{(1.16)}$ \\
                                                & BSM-MagLS & $5.08_{(0.53)}$ & $4.96_{(0.61)}$ & $4.81_{(0.72)}$  \\
                                                & BSM-iMagLS & $6.51_{(0.76)}$ & $6.53_{(0.77)}$ & $6.18_{(0.91)}$  \\
                                                & eMagLS & $7.21_{(0.87)}$ & $7.3_{(0.89)}$ & $6.77_{(1.21)}$  \\
        \hline
    \end{tabular}
    }
\end{table}

\subsubsection{Performance Under Reverberate conditions}\label{ch5_BRIR}

This section evaluates the performance of the proposed method under reverberant conditions within a simulated room environment. Figure~\ref{fig:BRIRS_analysis} presents the BSD results for each of the evaluated methods, using the EasyCom microphone array and the KU100 HRTF in the medium-sized room scenario. In the figure, the lines represent the mean BSD values, while the shaded regions indicate one standard deviation, both averaged across 100 randomly placed sources within the room.

The results reveal that all methods yield similarly low BSD values up to approximately \(1.5\,\text{kHz}\). This is expected, as all approaches share the same filter coefficients in this low-frequency range. Among the methods, BSM-LS exhibits the highest BSD values at higher frequencies, consistent with the known high-frequency roll-off phenomenon inherent in Least Squares optimization. This spectral roll-off leads to more pronounced coloration artifacts, especially in the upper frequency range. Above \(1.5\,\text{kHz}\), BSM-MagLS achieves the lowest BSD, with eMagLS demonstrating nearly identical performance up to \(10\,\text{kHz}\) and only a marginally increased distortion beyond that point. As anticipated, BSM-iMagLS remains closely aligned with the magnitude-optimized BSM-MagLS across the entire spectrum, in line with the trends observed under anechoic conditions.

A summary of the LSD results for all methods, across all evaluated microphone arrays and room sizes, is provided in Table~\ref{tab:metrics_summary_BRIRs}. The LSD values generally follow the same pattern observed in the medium room case shown in Figure~\ref{fig:BRIRS_analysis}, with only minor variations across different room sizes. These results suggest that the performance of all methods remains robust under varying levels of reverberation, showcasing the robustness to different acoustic conditions of these methods.

%% file: sections/Ch6_v3.tex
\section{Listening Experiment}
The perceptual benefits of the proposed BSM-iMagLS method were evaluated through a listening test. This test aimed to demonstrate the improvements observed in the simulated study by providing subjective assessments of BSM-iMagLS performance compared to BSM-MagLS and BSM-LS.

\subsection{Setup}
The microphone array signals were generated in a simulated acoustic environment consisting of a rectangular room with dimensions $[10 \times 6 \times 3]$m. The acoustic simulation employed the image method~\cite{allen1979image}, implemented in MATLAB (2023a). The room had a reverberation time of $T_{60}=0.48$sec, a reflection density of $30000$ reflections per second, a critical distance of $r_{\text{cd}}=1.1$m, and a maximum reflection order of $44$. A single omnidirectional point source was placed at $(x,y,z)= (5,3.7,1.7)$m. Two types of audio source stimuli were used: an English female speaker and castanets. The female speaker represents typical content for binaural reproduction applications, while the castanets were chosen due to their high-frequency content and strong transients. The microphone array signals $\mathbf{x}(k)$ were computed using the EasyCom ATF, with the array center located at $(4,2,1.7)$m, $2$m away from the source at a $+60^{\circ}$ azimuth angle and $+0^\circ$ elevation angle.

As these methods are not signal-dependent, BSM coefficients were calculated separately for each method, independent of the room simulation. All methods used the KU100 HRTF as $\mathbf{h}^{l,r}$ and the EasyCom ATF as $\mathbf{V}$. For head-tracking, BSM coefficients were precomputed and saved for each head orientation, by counter-rotating the HRTF in the opposite direction of head rotation. 

BSM coefficients were computed using three methods. The first method, BSM-LS, served as a baseline and was calculated using Eq.~\eqref{BSM-MSE} across the entire frequency range, assuming a high SNR of $+20$dB. The second method, BSM-MagLS, representing the state-of-the-art, was calculated using Eq.~\eqref{BSM-MLS} with a cutoff frequency of $1.5$kHz. Finally, the third method, BSM-iMagLS, which is the proposed approach, was calculated as described in Sec.~\ref{sec:3}, using the same training hyperparameters outlined in Sec.~\ref{ch5a}. 


The binaural signals for BSM-LS, BSM-MagLS, and BSM-iMagLS were computed using Eq.~\eqref{BSM_rendering}, saved for each head orientation, and then convolved with the stimuli for playback based on head-tracking readings. A reference signal was also rendered using high-order Ambisonics (order $35$) binaural rendering for the same acoustic environment and HRTF. All signals were convolved with matching headphone compensation filters from the Cologne database~\cite{bernschutz2013spherical}, measured on the Neumann KU100 dummy head. Loudness equalization across all test signals was performed using the algorithm specified in Recommendation ITU-R BS.1770-2~\cite{series2011algorithms}, which estimates perceived loudness based on a K-weighted measurement of audio signal energy. All signals were adjusted to match the loudness of the quietest stimulus. Participants listened to the signals using AKG K702 headphones with an external head-mounted head-tracker in a quiet test room environment.

 \begin{figure}
    \centering
    \includegraphics[width=0.45\textwidth,trim={2.0cm 0.0cm 3.0cm 1.0cm},clip]{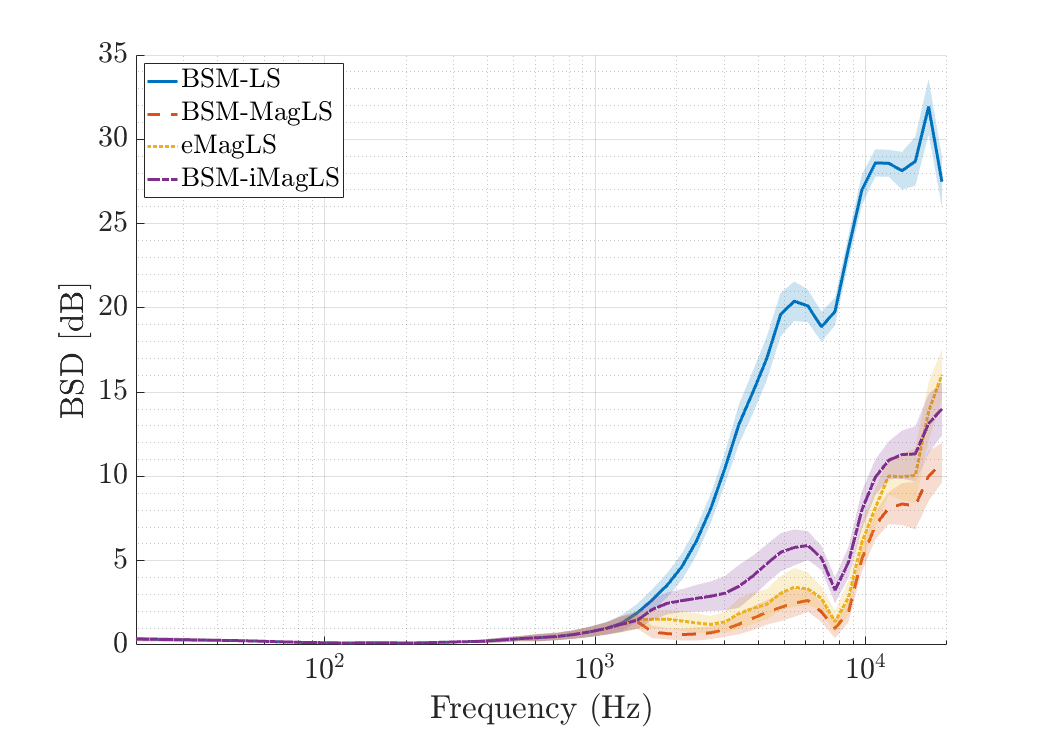}
    \caption{Reverberant room BRIR analysis, performed on KU100 and the measured EasyCom array. The figure shows the BSD for center frequency on the ERB scale from $20$Hz to $20$kHz. The curves depict mean and standard deviation with regards to the source location in the medium size room.} 
    \label{fig:BRIRS_analysis}
\end{figure}

\subsection{Methodology}
The listening tests followed the MUltiple Stimuli with Hidden Reference and Anchor (MUSHRA) protocol~\cite{series2014method}. Participants were asked to rate the spatial and timbre qualities of both stimuli, resulting in four MUSHRA screens, each containing four signals: a reference and three BSM methods. Of these, the LS can be considered as the lower anchor. Fourteen participants (eleven male and three female in the age range of $24-37$ years) without known hearing impairments, all familiar with spatial listening tests, participated in the study. The test was conducted in a quiet environment, and included a training phase and a signal familiarization phase. The training phase acquainted participants with the test equipment and rating scale, while the familiarization phase allowed participants to listen to all test signals. Participants were encouraged to rotate their head while listening and to compare between signals in various look directions. Participants rated the relative differences between the reference and test signals on a scale of $0$ to $100$, where $100$ represents no audible difference from the reference. Lower scores indicate larger perceptual differences. Participants assessed signals based on spatial quality (source localization and stability) and timbre quality (time-varying and spectral artifacts, overall tonal quality).

\subsection{Results}
\begin{figure}
    \centering
    \subfigure[Evaluation metric: Spatial quality]{\includegraphics[width=0.45\textwidth,trim={1.5cm 0.5cm 2cm 1cm},clip]{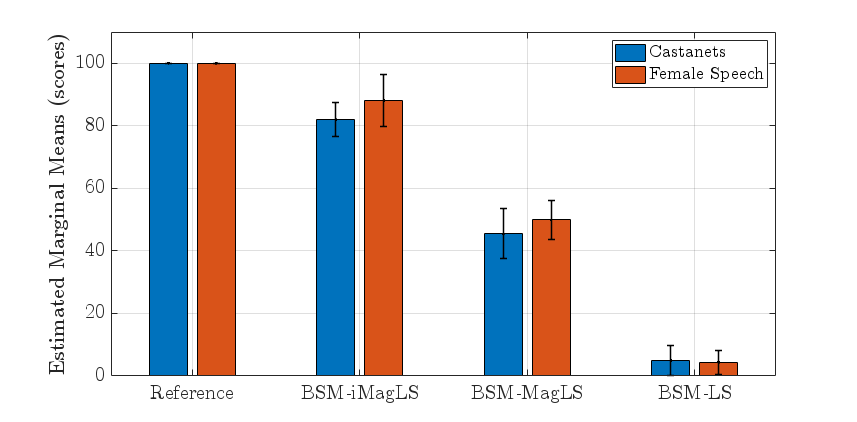}\label{fig:ch6_spacial}}
    \subfigure[Evaluation metric: Timbre quality]{\includegraphics[width=0.45\textwidth,trim={1.5cm 0.5cm 2cm 1cm},clip]{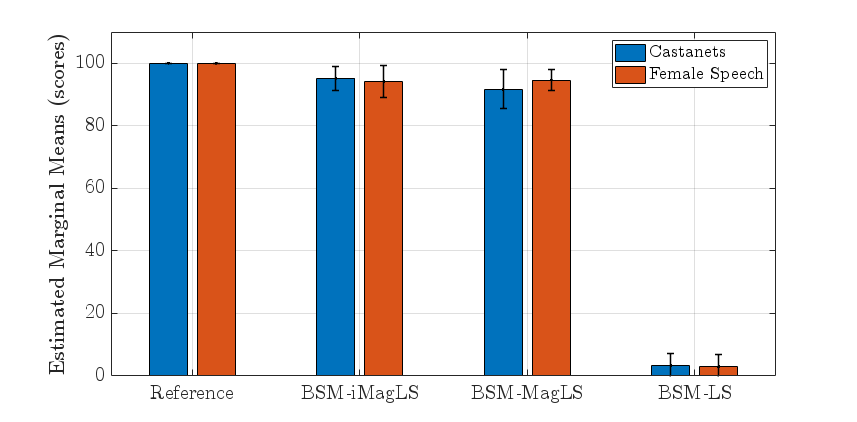}\label{fig:ch6_timbre}}
    \caption{Listening test results. Estimated marginal mean and $95\%$ confidence interval is shown for the BSM-LS, BSM-MagLS and the suggested BSM-iMagLS methods. Spatial quality results (top) and Timbre quality (bottom) are shown for the castanets and female speech in separate colors.}
    \label{fig:ch6_results}
\end{figure}

The participants' scores were analyzed using non-parametric repeated measures tests due to violations of normality~\cite{field2024discovering} rather then using a Repeated Measures Analysis of Variance~\cite{keselman2001analysis}. The within-subject factors were: Method (Reference, BSM-LS, BSM-MagLS, BSM-iMagLS), Stimulus (speaker, castanets), and Attribute (timbre, spatial). The estimated marginal medians and $95\%$ confidence intervals for the interactions are shown in Fig.~\ref{fig:ch6_results}.

Separate Friedman tests revealed statistically significant main effects of Method for both spatial and timbre quality scores. For spatial quality, $\chi^2(3) = 41.72$, $p < .001$, with a very large effect size ($W = .993$). For timbre quality, $\chi^2(3) = 34.54$, $p < .001$, also with a large effect size ($W = .823$). A Wilcoxon signed-rank test showed a significant main effect of Attribute, $Z = -3.30$, $p < .001$, with a large effect size ($r = .881$), indicating a difference in ratings for the attributes across methods and stimuli. A significant Method~×~Attribute interaction was also found, $\chi^2(3) = 31.19$, $p < .001$, with a large effect size ($W = .743$), indicating that the relative ranking of the methods differed depending on the evaluated attribute. In contrast, no significant main effect of Stimulus was observed, $Z = -1.19$, $p = .233$, $r = .319$, nor any other interactions between factors.

To further investigate the dependence on Method and Attribute, non-parametric post-hoc analyses were conducted using Wilcoxon signed-rank tests with Bonferroni correction~\cite{field2024discovering}.

\subsubsection{Attribute: Spatial Quality}
For the spatial quality attribute, statistically significant differences were found between all methods after Bonferroni correction ($\alpha = .0083$). A significant median difference ($p = .001$, $r = .85$) of $15$ points was observed between the Reference (median = $100$) and BSM-iMagLS. Additionally, large differences were observed between the Reference and BSM-MagLS ($p < .001$, $r = .88$, $52.3$ points), and BSM-LS ($p < .001$, $r = .91$, $95.3$ points). Differences between BSM-iMagLS and BSM-MagLS ($p < .001$, $r = .88$, $37.3$ points), and BSM-iMagLS and BSM-LS ($p < .001$, $r = .88$, $80.3$ points) were also significant. Lastly, the difference between BSM-MagLS and BSM-LS was statistically significant ($p < .001$, $r = .88$, $43$ points), confirming a consistent ranking among methods. These results suggest that the participants favored BSM-iMagLS over the other methods in terms of spatial quality, although it remained distinguishable from the reference.

\subsubsection{Attribute: Timbre Quality}
For timbre quality, a statistically significant difference was observed between the Reference and BSM-LS ($p < .001$, $r = .92$, $96.8$ points), and BSM-MagLS ($p = .008$, $r = .71$, $6.8$ points), exceeding the Bonferroni-corrected threshold ($\alpha = .0167$). The difference between the Reference and BSM-iMagLS was marginally non-significant ($p = .017$, $r = .64$, $5.3$ points), yet close enough to warrant interpretive consideration. Thus, it can be concluded that while there is a small difference in timbre between the reference and BSM-iMagLS and between the reference and BSM-MagLS, this difference is minimal, indicating that the timbre quality of both methods is comparable to the reference signal. As expected, BSM-LS performed poorly.

These results align with the objective evaluation in Sec.~\ref{ch5}, which similarly indicated that the proposed BSM-iMagLS method improves spatial attributes by reducing ILD error in the resulting binaural signals compared to BSM-MagLS and BSM-LS, while maintaining comparable spectral quality to the magnitude-optimized BSM-MagLS solution. Auralizations examples of BSM-iMagLS renderings are available as supplemental material~\cite{berebi_2025_15260730}.

%% file: sections/Ch7.tex
\section{Conclusion}

This paper introduced BSM with ILD and MagLS (BSM-iMagLS), a novel binaural rendering method designed for arbitrary microphone arrays, with a particular focus on head-mounted arrays. BSM-iMagLS enhances the previously proposed BSM method by incorporating ILD optimization. Through a DNN-based joint optimization framework, BSM-iMagLS significantly improves spatial accuracy, which is essential for effective binaural signal reproduction using wearable microphone arrays.

Our evaluation demonstrated that BSM-iMagLS consistently reduces ILD errors across various ATFs, HRTFs, and head orientations, while maintaining magnitude errors comparable to existing MagLS solutions. These improvements are especially pronounced in scenarios where horizontal localization is critical, thereby validating the perceptual benefits of the proposed method through both objective metrics and listening experiments.

The findings of this study emphasize the importance of balancing magnitude and ILD accuracy to enhance spatial perception in binaural audio reproduction. Consequently, BSM-iMagLS emerges as a more accurate and perceptually effective binaural rendering technique, particularly suited for the constraints of head-mounted and wearable microphone arrays.

Future work could include a more in-depth evaluation of the iMagLS method. This could involve a comparison with signal-dependent parametric methods to better understand the relative strengths and limitations. Testing in more challenging conditions, including complex acoustic scenes with multiple simultaneous sources, dynamic environments with moving sources, and recordings captured in real-world rather than simulated settings. In addition, the joint optimization framework could benefit from the inclusion of specific perceptually motivated metrics. These could include interaural time differences to support horizontal localization, pinna-related spectral cues to enhance localization in the polar dimension, and interaural coherence to improve the sense of envelopment and externalization. These extensions could broaden the applicability of BSM-iMagLS and refine its performance across a wider range of use cases.

%% file: bios.tex
\begin{IEEEbiography}[{\includegraphics[width=1in,height=1.25in,clip,keepaspectratio]{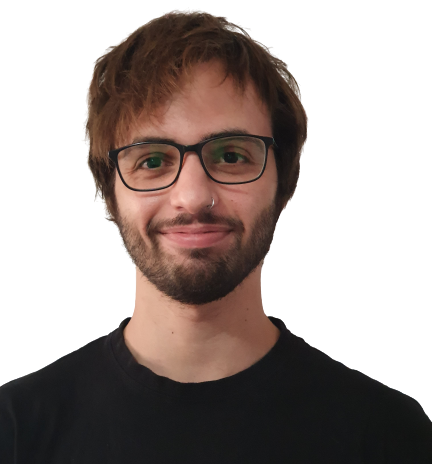}}]{Or Berebi}
received the B.Sc and M.Sc degrees in electrical and computer engineering in 2019 and 2022, respectively, from Ben-Gurion University of the Negev, Beer-Sheva, Israel, where he is currently working towards the Ph.D. degree. His current research interests focus on improving spatial audio perception for low order binaural reproduction.
\end{IEEEbiography}

\begin{IEEEbiography}[{\includegraphics[width=1in,height=1.25in,clip,keepaspectratio]{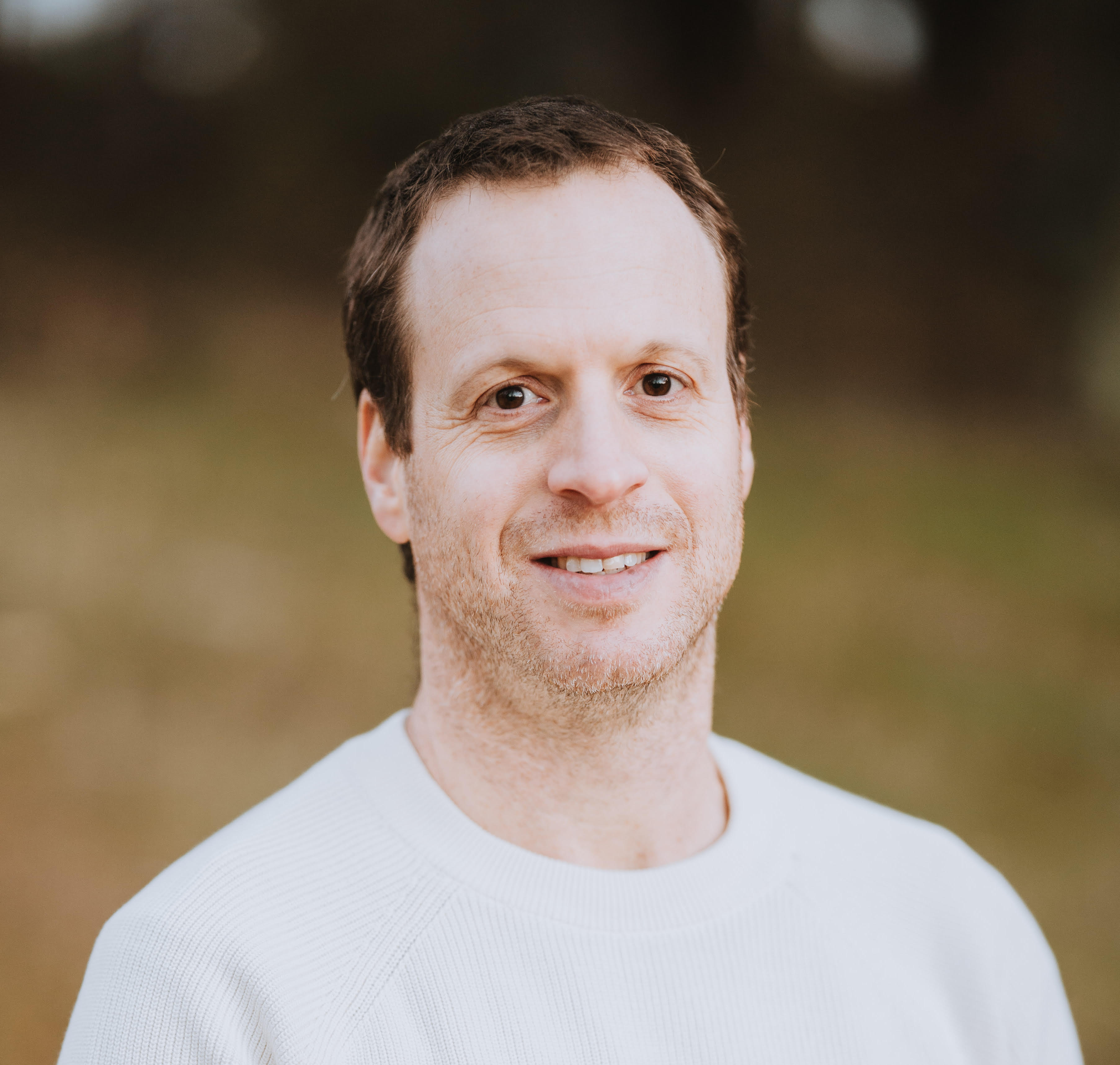}}]{Zamir Ben-Hur}
is currently a research scientist at Meta Reality Labs Research, working on spatial audio technologies. He received the B.Sc. (summa cum laude), M.Sc. and Ph.D. degrees in electrical and computer engineering in 2015, 2017 and 2020 respectively, from Ben-Gurion University of the Negev, Beer-Sheva, Israel. His research interests include spatial audio signal processing for binaural reproduction with improved spatial perception.
\end{IEEEbiography}

\begin{IEEEbiography}[{\includegraphics[width=1in,height=1.25in,clip,keepaspectratio]{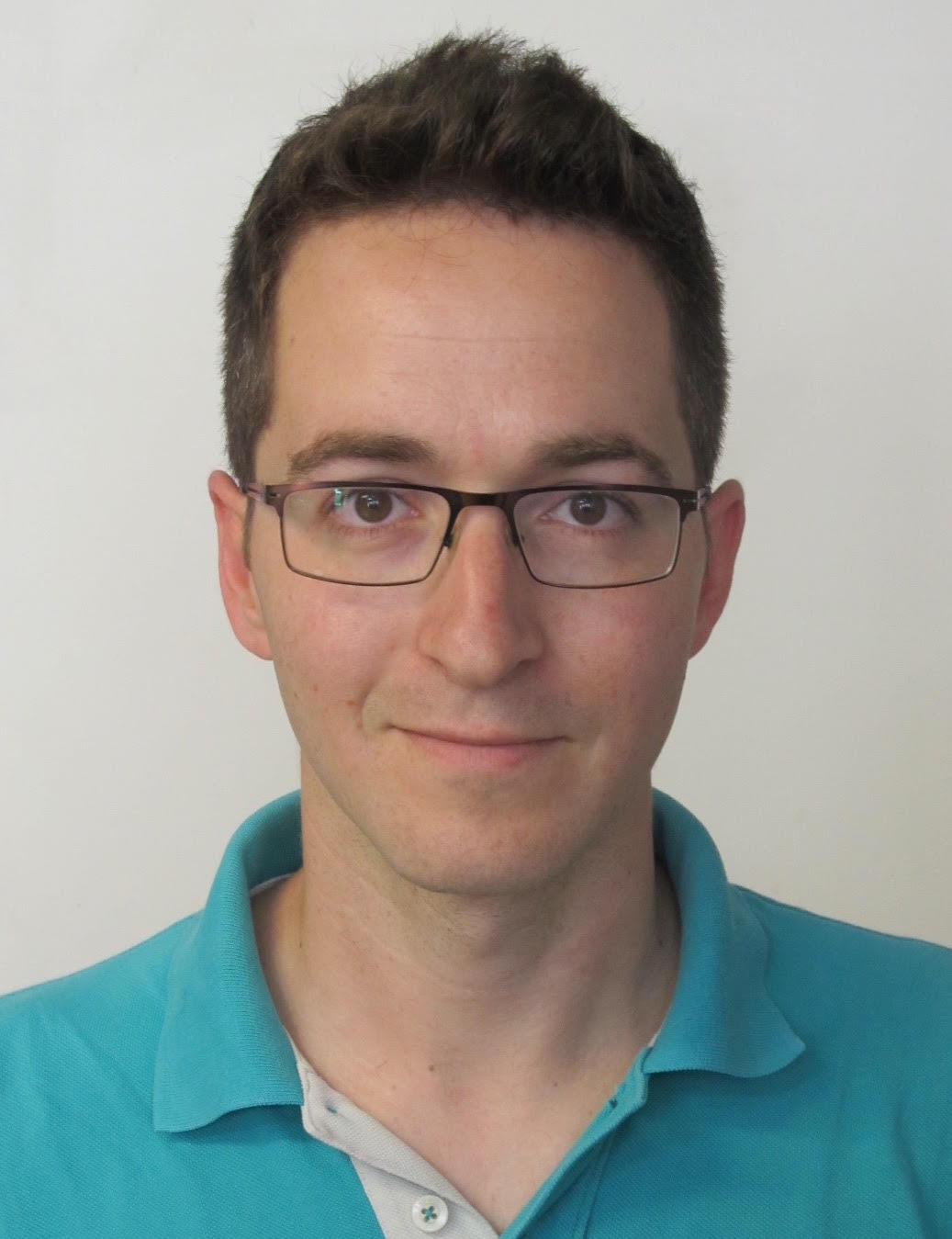}}]{David Lou Alon}
is a Research Scientist at Meta Reality Labs Research, investigating spatial audio technologies. He received his Ph.D. in electrical engineering from Ben Gurion University (Israel, 2017) in the field of spherical microphone array processing. His research areas include head-related transfer functions, spatial audio capture, binaural reproduction, and headphone equalization for VR and AR application.
\end{IEEEbiography}

\begin{IEEEbiography}[{\includegraphics[width=1in,height=1.25in,clip,keepaspectratio]{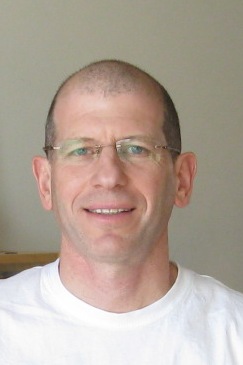}}]{Boaz Rafaely}
(SM’01) received the B.Sc. degree (cum laude) in electrical engineering from Ben-Gurion University, Beer-Sheva, Israel, in 1986; the M.Sc. degree in biomedical engineering from Tel-Aviv University, Israel, in 1994; and the Ph.D. degree from the Institute of Sound and Vibration Research (ISVR), Southampton University, U.K., in 1997. 

At the ISVR, he was appointed Lecturer in 1997 and Senior Lecturer in 2001, working on active control of sound and acoustic signal processing. In 2002, he spent six months as a Visiting Scientist at the Sensory Communication Group, Research Laboratory of Electronics, Massachusetts Institute of Technology (MIT), Cambridge, investigating speech enhancement for hearing aids. He then joined the Department of Electrical and Computer Engineering at Ben-Gurion University as a Senior Lecturer in 2003, and appointed Associate Professor in 2010, and Professor in 2013. He is currently heading the acoustics laboratory, investigating methods for audio signal processing and spatial audio. During 2010-2014 he has served as an associate editor for IEEE Transactions on Audio, Speech and Language Processing, and during 2013-2018 as a member of the IEEE Audio and Acoustic Signal Processing Technical Committee. He also served as an associate editor for IEEE Signal Processing Letters during 2015-2019, IET Signal Processing during 2016-2019, and currently for Acta Acustica. During 2013-2016 he has served as the chair of the Israeli Acoustical Association, and is currently chairing the Technical Committee on Audio Signal Processing in the European Acoustical Association. Prof. Rafaely was awarded the British Council’s Clore Foundation Scholarship.
\end{IEEEbiography}